\documentclass[reprint,amsmath,amssymb,aps,prc,superscriptaddress,twocolumn,nofootinbib]{revtex4-2}

\usepackage{graphicx}
\usepackage{dcolumn}
\usepackage{bm}
\usepackage[colorlinks=true,urlcolor=blue]{hyperref}
\usepackage{amsmath}
\usepackage{amssymb}
\usepackage{color}

\begin{document}
	\title{On the extraction of fission mode properties from fragment mass distributions}

	\author{Patrick McGlynn}
	\affiliation{Fundamental and Theoretical Physics, \& Nuclear Physics and Accelerator Applications, \\ Australian National University, Acton ACT 2601 Australia }
	\affiliation{Facility for Rare Isotope Beams, Michigan State University,\\ 640 South Shaw Lane, East Lansing MI 48824 USA}
	\author{C\'edric Simenel}
	\affiliation{Fundamental and Theoretical Physics, \& Nuclear Physics and Accelerator Applications, \\ Australian National University, Acton ACT 2601 Australia }
	\date{\today}
	
	\begin{abstract}
	\noindent\textbf{Background:} Fission modes are typically characterised by fragment mass and total kinetic energy centroids, around which a distribution of these variables is observed. These distributions are usually fitted with Gaussian functions.
	
	\noindent\textbf{Purpose:} To investigate how the properties of these ``Gaussian fission modes'' compare with underlying ``theoretical fission modes'' defined from the potential energy surface of the fissioning nuclei.

	\noindent\textbf{Methods:} A simple approach, inspired by the scission point model, is introduced 
	to investigate the impact of anharmonicity of the potential along the scission line on Gaussian mode properties. 
	This approach is also used to evaluate an ``effective potential'' from the yields. 

	\noindent\textbf{Results:} Several Gaussian functions are usually required to fit yields from non-harmonic potentials associated with a unique theoretical mode. 
 ``Effective fission modes'', defined from the wells of the effective potentials, can be very different to the Gaussian modes. 
For instance, the S1 and S2 Brosa modes contribute to the same effective potential well at scission.
			
	\noindent\textbf{Conclusions:} Although Gaussian fits have proven useful to identify the role of shell effects in fission, they are not always associated with fission modes originating from valleys in the potential energy surfaces. Instead, properties of fission modes are better investigated with effective potentials extracted from the yields, preferably with a broad range of excitation energies.
		
	\end{abstract}

	\maketitle
	
	\section{Introduction}

	Nuclear fission allows for a sensitive dynamic probe of nuclei and  the behaviour of nucleons within them.	Even the earliest measurements of the mass and charge yields of fission fragments \cite{Meitner1939,Grummitt1946} identified a distribution of masses, with some nuclides tending to fission asymmetrically, and others symmetrically. Labelling these as fission modes swiftly followed \cite{Turkevich1951}.
	The symmetric and asymmetric  modes exhibit total kinetic energy (TKE) distributions characteristic to the charge asymmetry  as well as the elongation of the pre-scission nucleus  \cite{Britt1963}. 
	 The origin of these modes is nowadays interpreted in terms of shell effects in the fissioning nucleus \cite{gustafsson1971,ichikawa2019,cwiok1994,bernard2023} as well as in the fragments \cite{Wilkins1976,sadhukhan2016,scamps2018,scamps2019,mahata2022,bernard2023}. 
	In particular, Brosa et al \cite{Brosa1990} tied the fission modes to particular pre-scission shapes. Although they approximated the distribution of observables from each mode as being a normal or Gaussian distribution, they also  asserted that they ``do not consider the Gaussian to be more than an approximate	representation of the yield from one fission channel" and note that experimental evidence for the existence of distinct modes requires detailed measurement of the change in yields with small changes in excitation energy and neutron number, to reveal the bifurcations between these channels. 
	
	In the 21st century, continual improvements in accelerator and detector technology have allowed for more and more measurements of fission fragment mass and TKE, revealing a variety of fission modes. Many publications do not explicitly fit particular modes to their data \cite{Caamano2015,Caamano2013,Schmidt2000,Schmidt2001,Fregeau2016,Chatillon2019,Nishio2008,Chatillon2023,Jhingan2022,Fernandez2023}, however those that do almost invariably fit the fragment mass (or charge) distributions using a collection of Gaussians \cite{Simutkin2014,Kozulin2022,Leguillon2016a,Swinton-Bland2020b,Swinton-Bland2023a,Banerjee2022,Nag2021,Prasad2020a,Berriman2022}. Assuming that a symmetric mode leads to a Gaussian distribution centred around symmetric mass split, and that each asymmetric  mode is associated with a pair of Gaussians (centred around the most likely heavy and light fragments for this mode), leads to the interpretation of multiple modes based on the quality of these fits. However, the underlying assumption of the Gaussian mode as a represenatation of a fission mode is unquestioned, despite the fact that theoretical models like the scission point model \cite{Wilkins1976} or the time-dependent generator coordinate method (TDGCM) \cite{Goutte2005,Schunck2016} usually do not predict yields with Gaussian shapes. 
Nevertheless, Gaussian fits of yield distributions have been used to identify the role of shell effects through the stability of Gaussian centroids with respect to changes in excitation energy and fissioning nucleus. However, it is not clear how these ``Gaussian modes'' translate into features of the underlying potential energy surface (PES). 
Indeed,  the yield distributions could in principle be used to constrain the shape of the PES near to the scission line \cite{Bartel2014,Randrup2011b,McDonnell2014,Schunck2014,Kostryukov2021,Zdeb2021,Okada2023a}.

The purpose of this work is to evaluate the relationship between fission mode properties extracted from Gaussian fits of fragment mass or charge distributions, and the properties of the potential energy of the fissioning system near scission, such as fission valleys in PES. 
In particular, we introduce a simple expression to extract an ``effective potential energy'' that can be used to compare experimental yields with theoretical PES properties near scission, provided that a broad range of excitation energies is considered. We also highlight the inconsistence of interpreting Gaussian centroids as evidence of fission modes.

Expressions used to parametrise fragment mass asymmetries and yields, as well as the various definitions of fission modes used in this work, are provided in Sec.~\ref{sec:def}.
Section~\ref{sec:VtoY} discusses the impact of the potentials on the yields and their Gaussian fits properties. 
Effective potentials are extracted from fits of experimental data, and their dependence with excitation energies are analysed in Sec.~\ref{sec:YtoV}. 
Conclusions are drawn in Sec.~\ref{sec:conc}.

	\section{Model}

\subsection{Parametrisation of fragments mass asymmetry \label{sec:def}}

	For simplicity, we assume that there is a single possible TKE for a given mass split.
	 This allows us to parameterise yields with one coordinate associated with mass repartition between fragments. 
 We define the coordinate $x$ to compare the mass ratio of the fragments in a way that allows easy fitting and comparison, and is unitless:
	\begin{equation}
		x=4\frac{A_1-A_{T}/2}{A_{T}},
	\end{equation}
	where $A_1$ is the mass number of one fragment and  $A_{T}$ is the  mass number of the compound nucleus.
	By definition, $x$ ranges from $-2$ to $2$, however the region of interest is generally contained within $[-1,1]$.
Note that, under the  unchanged charge density assumption, $x$ can also be used to compare charge ratio of the fragments. 

\subsection{Parametrisation of yield distribution}

Inspired by the scission point model \cite{Wilkins1976} (see also  \cite{Pasca2016,Pasca2018,Pasca2020,Pasca2021,Carjan2017,Carjan2019,Carjan2015,Pasca2023}), we write the yield as 
\begin{equation}
Y(x)= e^{-\frac{V(x)}{T}}.\label{eq:yield}
\end{equation}
In the scission point model, i.e., under the assumption of quasiequilibrium among collective degrees of freedom near the scission point, $V(x)$ is 
the potential energy of a shape associated with $x$ immediately before scission, while $T$ is the temperature of the collective system.
Equation~(\ref{eq:yield}) will also be used as a way to parametrise the yield, in which case $V(x)$ will be treated as an ``effective potential'' used as a proxy for the actual potential along the scission line.
In particular, the effective potentials will be used to interpret the Gaussian fits of the yield in terms of potential shapes. The ``temperature'' $T$ is used to account for a dependence with the excitation energy $E^*$ of the form  $T=\sqrt{E^*/a}$, with $a=10$~MeV$^{-1}$.
The yield normalisation only leads to a constant shift of $V(x)$ and is irrelevant. 
Other models which consider random walks \cite{Moller2015a,Moller2012} or diffusion paths \cite{Aritomo2013,Aritomo2014,Liu2021,Kostryukov2021,Schmitt2019} or even coherent wavepacket evolution \cite{Regnier2019,Verriere2020a,Zhao2022,Schunck2016,Younes2019} along a potential energy surface are not exactly equivalent to this simple picture. Nevertheless, in the end, the topography of the potential energy always plays a significant role in the resulting shape of the yield distribution.

\subsection{Definitions of fission modes}

The definition of a fission mode (or fission channel) is not necessarily the same from a theoretical or experimental perspective.
Experimentally, fission modes are associated with structures in the fragment mass, charge, and TKE distributions, usually fitted with Gaussian functions. 
In particular, different modes correspond to different Gaussian centroids.
Theoretically, fission modes are associated with valleys in the PES, or, in the context of the scission point model, minima in the scission line.
In this case, different modes correspond to different local minima separated by potential energy ridges or barriers. 

As the potential energy is in principle not an observable, the connection between both characterisations of fission modes is not straightforward. 
To facilitate this connection, we propose an alternative definition of fission modes based on the effective potential $V(x)$.
The latter can be evaluated directly from the yield distribution using Eq.~(\ref{eq:yield}).
The minima of $V(x)$ (which is a symmetric function) then indicate the degree of asymmetry of the fission modes. 
A symmetric mode is associated with a local minimum at $x=0$, while an asymmetric mode is associated with two local minima\footnote{In principle, the barrier between the minima should be at least of the order of the zero point energy.} at $x=\pm x_0\ne0$.  

In agreement with a quantum treatment that predicts Gaussian ground-states of harmonic oscillators, a symmetric Gaussian yield would lead to a harmonic effective potential  $V(x)\propto x^2$.
Similarly, at low enough excitation energy, one could expect asymmetric wells in $V(x)$ to produce a Gaussian yield centred around $\pm x_0$ due to the approximate quadratic shape of the potential near its minima at $\pm x_0$. 
As will be discussed in details in this work, however, yields that require several Gaussians for a reasonable fit do not necessarily lead to one minimum per Gaussian in the effective potential, suggesting that the theoretical and experimental definitions of fission modes are not equivalent. 

For clarity, let us summarise the three definitions of fission modes that are used in this work:
\begin{itemize}
\item {\it Gaussian modes} are associated with the characteristics of the Gaussian functions traditionally used to fit the yield distribution;
\item {\it theoretical modes} correspond to valleys in the PES, or in the context of the scission point model, minima of potential energy along the scission line;
\item minima in the effective potential $V(x)$ [see Eq.~(\ref{eq:yield})] are called {\it effective modes}. 
\end{itemize}
In Sec.~\ref{sec:VtoY}, we analyse the yields induced by potentials within a simple approach based on the scission point model, that is then used to compare theoretical, Gaussian, and effective modes. 
In Sec.~\ref{sec:YtoV}, Gaussian modes obtained from fits of experimental fission data are used to evaluate the effective potential, and then to compare Gaussian and effective modes. 

\section{From potential to yield \label{sec:VtoY}}

In this section we investigate the shape of yield distributions induced by simple non-harmonic potentials $V(x)$.
Here, $V(x)$ can be interpreted as  describing the potential energy along the scission line, rather than being an effective potential.
Following the scission point model, the yields are obtained from Eq.~(\ref{eq:yield}) assuming a quasiequilibrium among collective degrees of freedom near the scission point.
This approximation neglects pre-scission dynamics and does not account for the effect of, e.g., large barriers that could prevent the system to explore a particular fission valley. 
Nevertheless, this simple model is sufficient to illustrate the difference between mode characterisations. 

\subsection{Symmetric mode}

	\begin{figure}
		\includegraphics[width=0.45\textwidth]{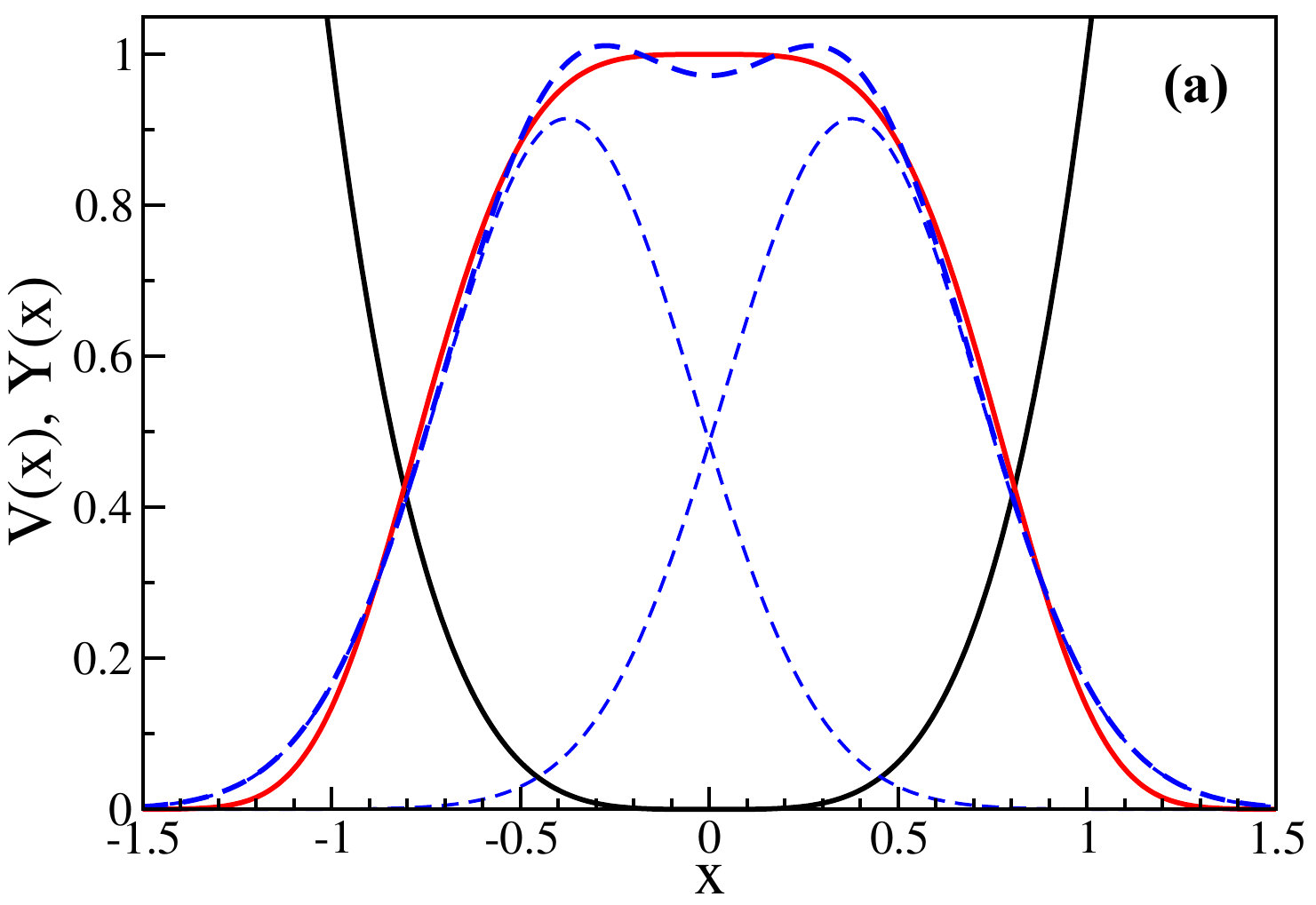}
		\includegraphics[width=0.45\textwidth]{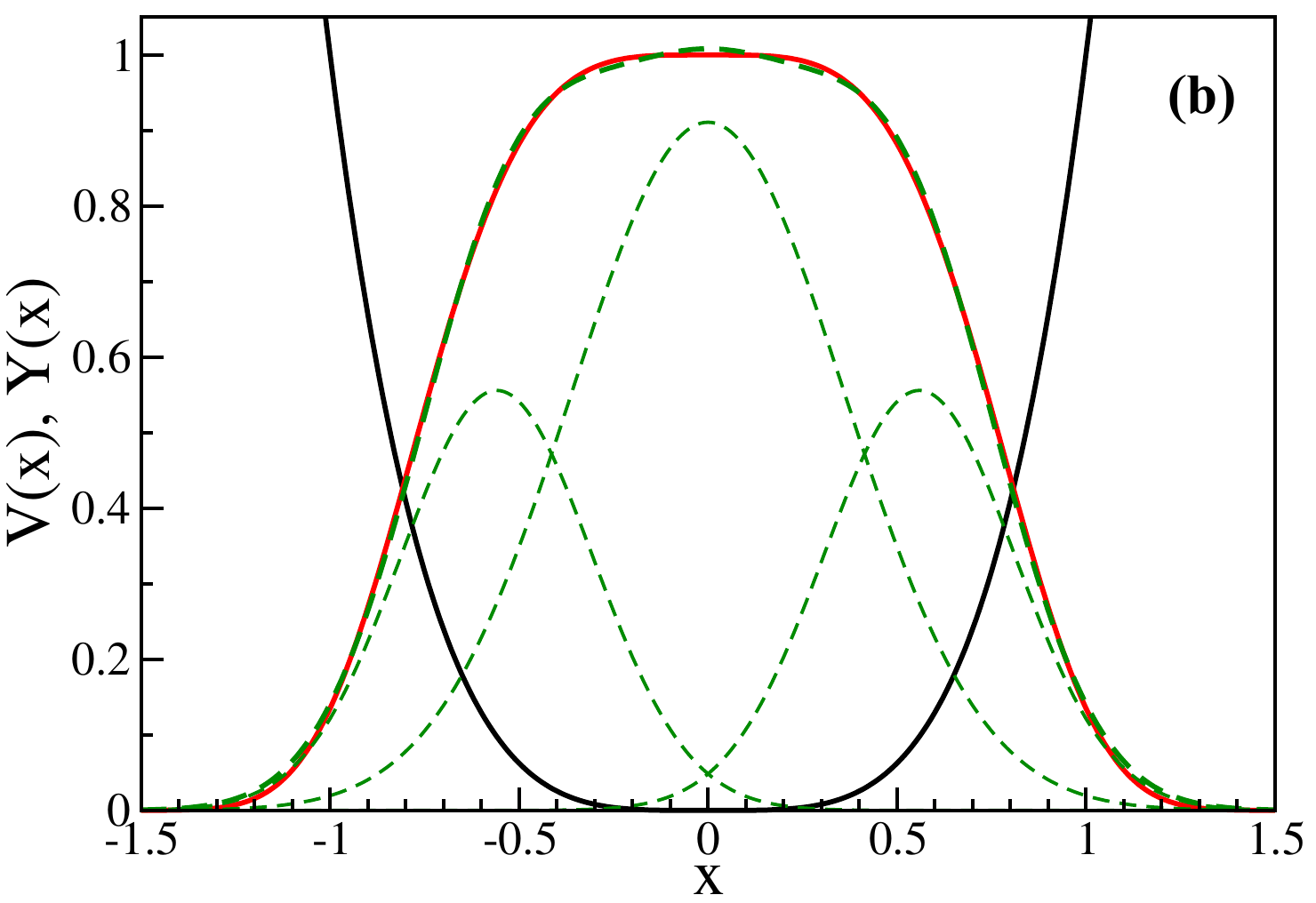}
		\caption{Quartic potential $V=x^4$ (black solid line) in MeV, and associated yield $Y(x)$ from Eq.~(\ref{eq:yield}) with $T=0.5$~MeV  (red solid line). (a) Two- and (b) three-Gaussian fits (thick dashed lines). Thin dashed lines show each individual Gaussian component.}
		\label{fig:quartic}
	\end{figure}

As mentioned in the previous section, a quadratic potential $V\propto x^2$ naturally leads to a yield with a symmetric peak (i.e., centred at $x=0$) and a Gaussian shape, in which case the Gaussian mode is a perfect representation of the theoretical mode. 
In a polynomial expansion of the potential, however, the quadratic term could also vanish, in which case one would expect a yield distribution deviating from a Gaussian shape. 
Consider, as an example, a quartic potential $V=x^4$ as represented in Fig.~\ref{fig:quartic}. The associated yield $Y(x)$ from Eq.~(\ref{eq:yield}) with $T=0.5$~MeV  is clearly not Gaussian. 
Figs.~\ref{fig:quartic} (a) and (b) also show two- and three-Gaussian fits to $Y(x)$. The fitted functions (up to an irrelevant overall normalisation) are 
\begin{align}
    Y(x)&=\left[(1-y)\left(e^{-\frac{(x-x_0)^2}{2\sigma_1^2}}+e^{-\frac{(x+x_0)^2}{2\sigma_1^2}}\right)+ye^{-\frac{(x)^2}{2\sigma_2^2}}\right]\nonumber
\end{align}
with $y=0$ in the two-Gaussian (one asymmetric mode) case [Fig.~\ref{fig:quartic}(a)] and $0<y\le1$ otherwise. The apparent quality of the fit is poor with two Gaussians, but gets much better with three Gaussians. Covariance analysis of a fit can be used to identify sources of overfitting. The symmetric covariance matrix $\text{cov}(x_i,x_j)$ contains information about the fitted parameters' uncertainty and correlation based on the data to which a model is fitted. In general, larger (compared with the scale of the parameter) on-diagonal elements suggest that the data cannot constrain a parameter well since it has a large variance. The strength of correlation is best viewed via the correlation matrix, in which covariances are normalised by the geometric mean of independent parameter variances.  Large off-diagonal elements of the correlation matrix suggest overfitting, as changing one parameter significantly affects the best-fit value for another parameter. Off-diagonal elements of the correlation matrix have a magnitude $\sim$1 in case of overfitting. Analysis of the covariance and correlation matrices for each fit (see Tab.~\ref{tab:Correlation}) suggests that while the two-Gaussian fit has largely uncorrelated parameters, the three-Gaussian fit has uniformly high correlation, implying overfitting. A fit of almost equal quality can be achieved by changing pairs of fit parameters in a way that cancels out, meaning that with this data, only a smaller number of linear combinations of these parameters are uniquely constrained. Any fit to a similar realistic dataset with three or more Gaussians should be cautiously interpreted especially when extracting information about mode positions, relative yields and widths; in practice the data are only able to constrain some subset of these.
\begin{table}
    \centering
\begin{tabular}{r|rr}
2-Gaussian&$\sigma_1$ & $x_0$ \\
\hline
$\sigma_1$& 2.3215e-05 & -1.4133e-07 \\
$x_0$& \textbf{-0.0092} & 1.0181e-05 \\
\end{tabular}

\begin{tabular}{r|rrrr}
3-Gaussian& $\sigma_1$ & $x_0$ & $y$ & $\sigma_2$ \\
\hline
$\sigma_1$&1.7674e-03 & 3.3744e-03 & -1.8522e-02 & -1.1459e-02 \\
$x_0$ & \textbf{0.9990} & 6.4560e-03 & -3.5448e-02 & -2.1933e-02 \\
$y$ & \textbf{-0.9979} & \textbf{-0.9993} & 1.9490e-01 & 1.2063e-01 \\
$\sigma_2$&\textbf{-0.9975} & \textbf{-0.9990} & \textbf{0.9999} & 7.4670e-02 \\
\end{tabular}
    \caption{On-diagonal and upper-right: elements of the covariance matrix for the two- and three-Gaussian fits respectively in Fig.~\ref{fig:quartic}. In bold below the diagonal: correlation matrix elements, defined as $C_{ij}=\text{cov}(x_i,x_j)/\sqrt{\text{cov}(x_i,x_i)\text{cov}(x_j,x_j)}$. The covariances are calculated on the assumption of a uniform uncertainty, with data points evaluated at intervals of $0.1$ in $x$ from -1.5 to 1.5.}
    \label{tab:Correlation}
\end{table}

	\begin{figure}
		\includegraphics[width=0.45\textwidth]{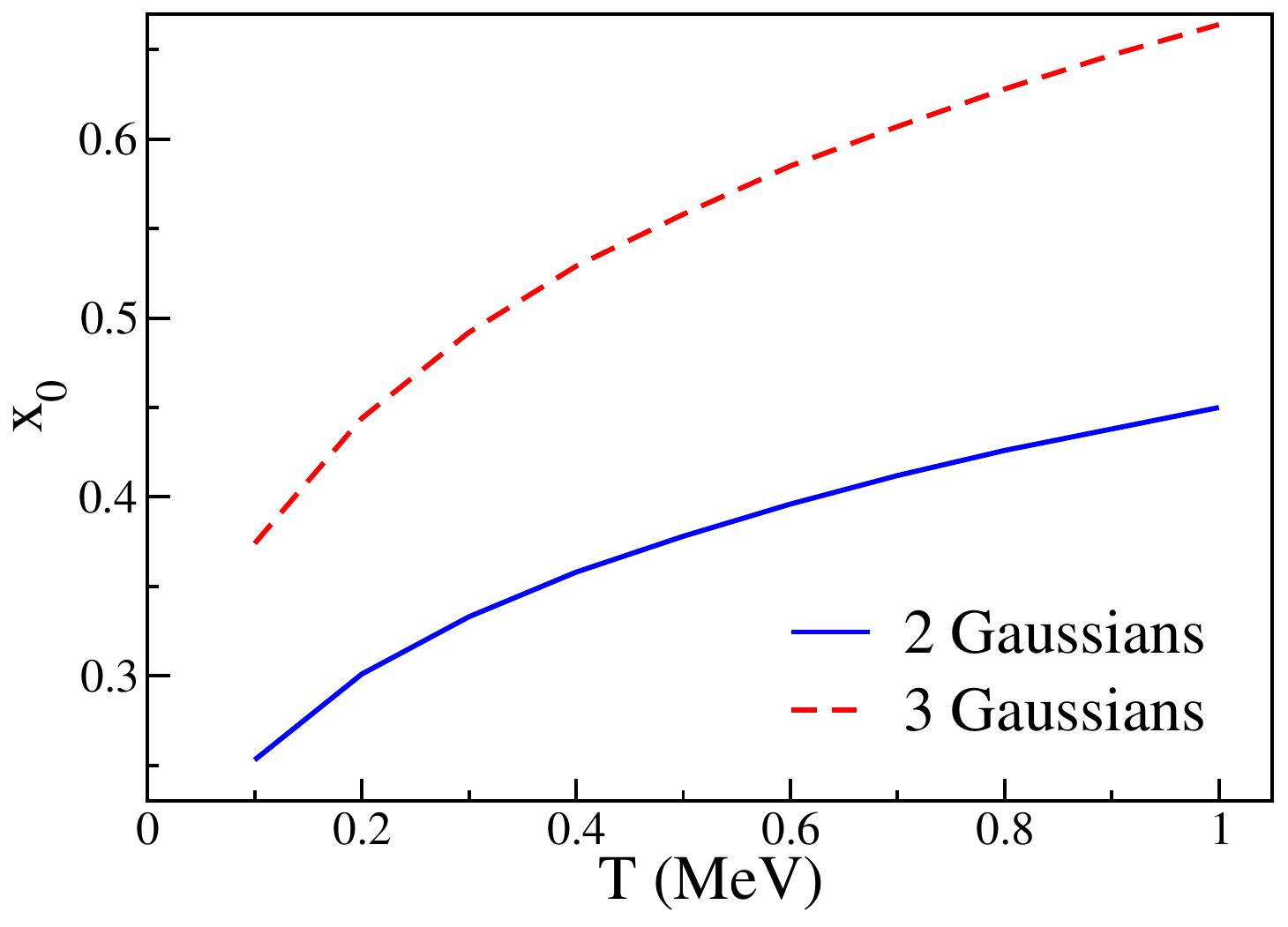}
		\caption{Centroids of the asymmetric Gaussian modes obtained from two- and three-Gaussian fits as a function of $T$ for a quartic potential.}
		\label{fig:sym_temp}
	\end{figure}

The centroids of the asymmetric Gaussian mode  vary significantly with excitation energy, as shown in Fig.~\ref{fig:sym_temp}, despite the fact that the underlying potential remains the same. 
Based on the better agreement obtained with the three-Gaussian fit, one might conclude that the system exhibits both a symmetric and an asymmetric Gaussian mode. 
However, the underlying physics is clear: there is only one symmetric theoretical mode in this system. The dependence of the centroid of the asymmetric Gaussian mode with excitation energy is also an indication that it is not associated with a theoretical mode. This latter property would not be apparent to a measurement conducted at a single energy.

	\begin{figure}
		\includegraphics[width=0.45\textwidth]{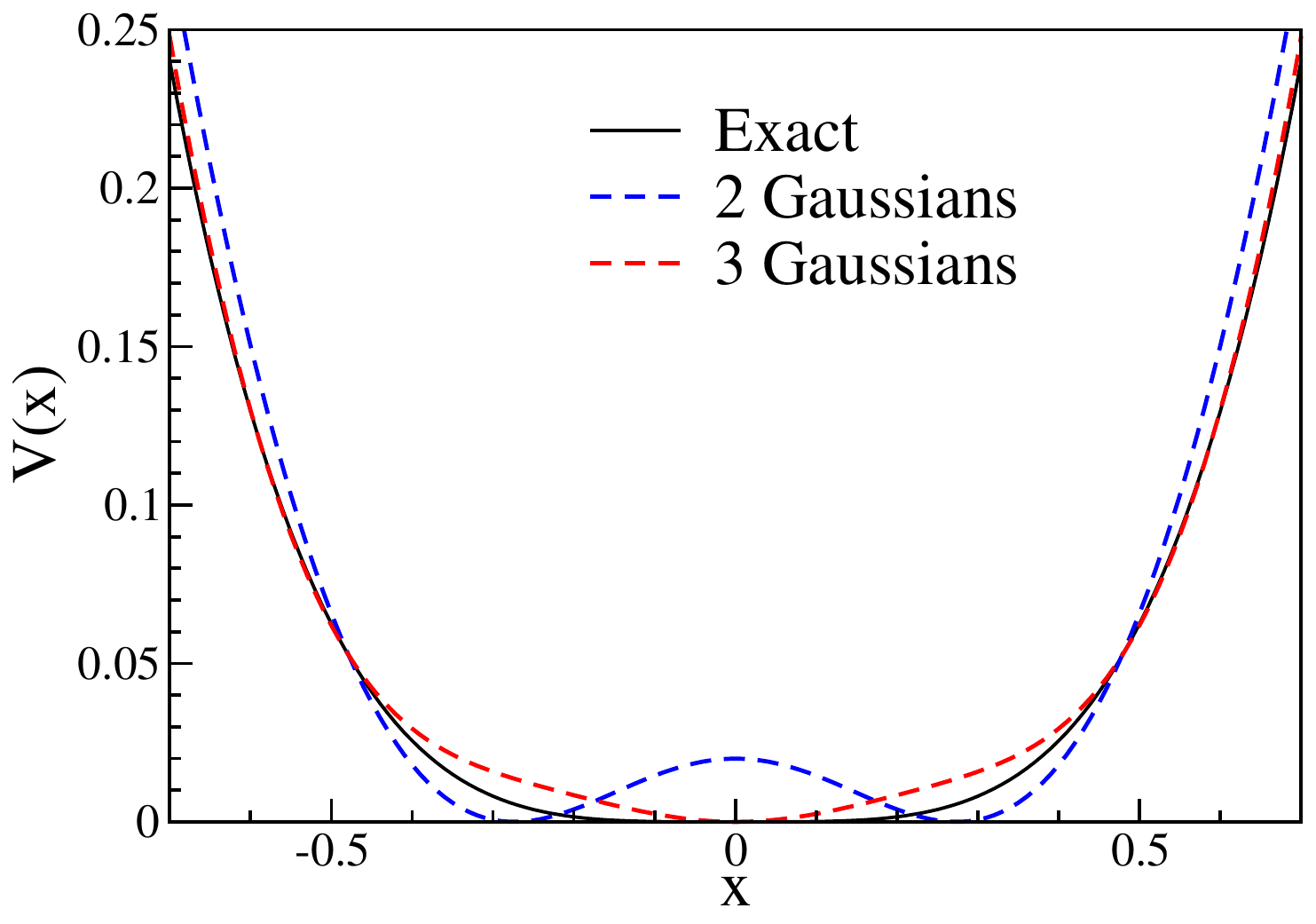}
		\caption{Comparison between the quartic potential (solid line) and the potentials extracted from the two (blue dashed line) and three (red dashed line) Gaussian fits in Fig.~\ref{fig:quartic} using Eq.~(\ref{eq:VG}).}
		\label{fig:sym_pot}
	\end{figure}

The Gaussian fits of the yield 
$$Y_{G}(x)=\sum_{n=1}^{N_G}y_ne^{-\frac{(x-x_n)^2}{2{\sigma_n}^2}}=Ce^{-\frac{V_G(x)}{T}},$$ 
with $N_G$ the number of Gaussians, 
 can also be used to evaluate the associated effective potential 
\begin{equation}
V_G(x)=-T\left[\ln\left(Y_G(x)\right)-\ln C\right],\label{eq:VG}
\end{equation}
where $C$ is a constant.
The resulting potentials are plotted in Fig.~\ref{fig:sym_pot}. 
Although the potential obtained from the two-Gaussian fit  leads to a spurious asymmetric effective mode, the three-Gaussian fit correctly predicts a potential with a unique symmetric effective mode. 
This demonstrates that, for Gaussian fits to be compared with theoretical modes, they should be used to extract an associated effective potential, as in Eq.~(\ref{eq:VG}). 
The data can then be said to contain a certain number of effective modes corresponding to minima of the effective potential, and not to the number of Gaussians used to make a good fit.
In the framework of the scission point model, theoretical and effective modes have similar characteristics, providing the fit of the yield is of good quality. 
In the present case, this requires a minimum of three Gaussians but still predicts a single effective mode. 
Further increasing the number of Gaussians in the fit would not change the conclusions as this would have a small impact on the effective potential. 

Of course, one is not required to use Gaussians at all. 
Indeed, the potential can be extracted directly, e.g., using a polynomial expansion $V(x)=\sum_{n=1}^{N_p}a_nx^{2n}$, where $N_p$ is the degree of the polynomial, and fitting the yield with 
\begin{equation}
Y(x)=C\exp\left(-\frac{1}{T}\sum_{n=1}^{N_p}a_nx^{2n}\right)\label{eq:Ypolyn}
\end{equation}
to extract the coefficients $a_n$. 
In doing so, there is not need to introduce Gaussian modes and one avoids the risk of over-interpreting their features in terms of physical properties. 

\subsection{Asymmetric mode}

	\begin{figure}
		\includegraphics[width=0.45\textwidth]{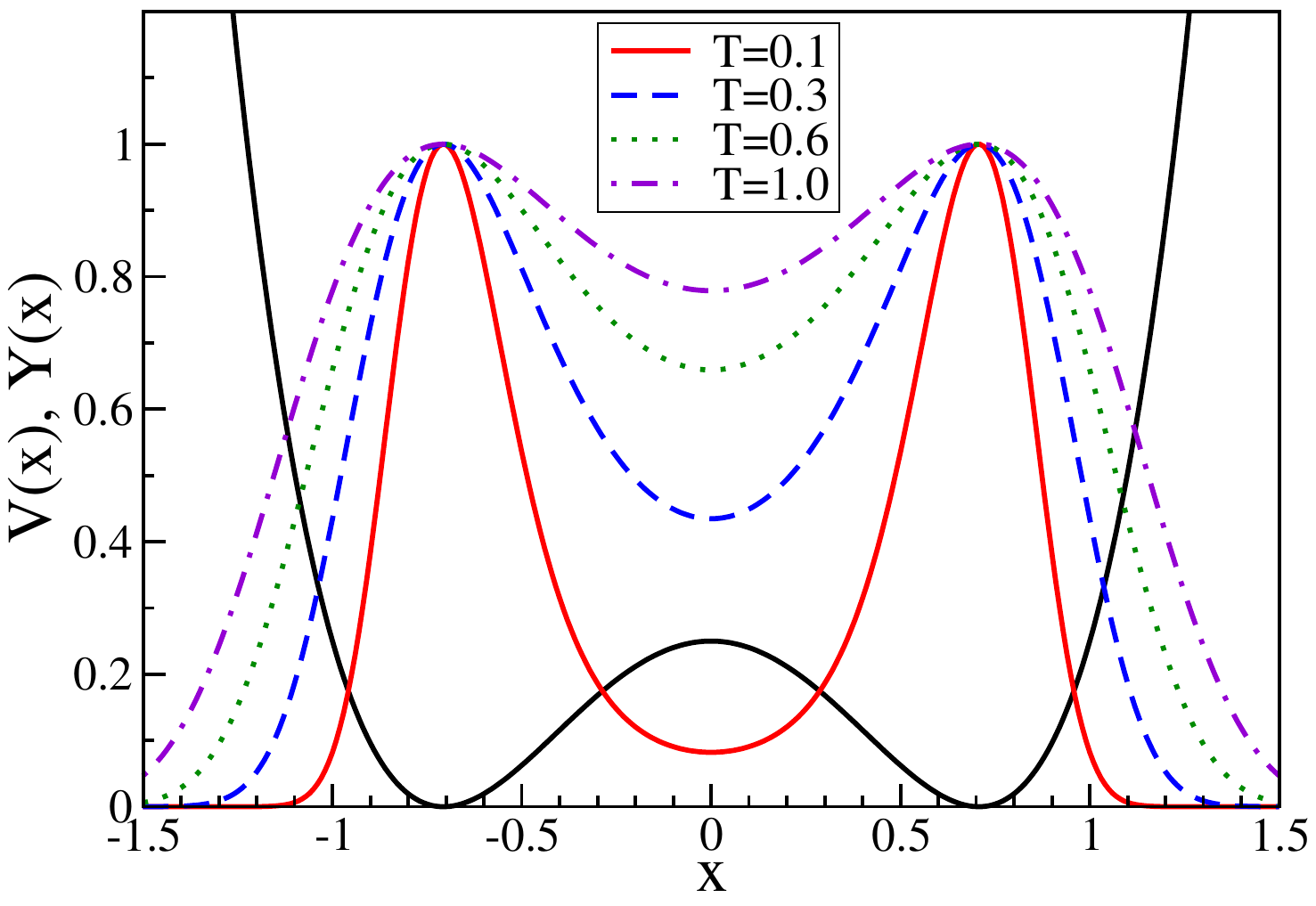}
		\caption{Two-well potential $V(x)=\frac{1}{4}-x^2+x^4$ (black solid line) in MeV, and yields $Y(x)=e^{-V(x)/T}$. Values of $T$ are in MeV. }
		\label{fig:asym_T}
	\end{figure}

Figure~\ref{fig:asym_T} shows an example of a two-well potential associated with one asymmetric theoretical mode with $x_0=1/\sqrt{2}$.
Yields obtained at various excitation energies are also shown. 
As expected, the lower the excitation energy, the closer the peaks are to a Gaussian shape. 
At higher $T$, however, the increase of the yield at symmetry would require an additional symmetric Gaussian mode, despite the fact that the potential does not exhibit any symmetric theoretical mode. 

	\begin{figure}
		\includegraphics[width=0.45\textwidth]{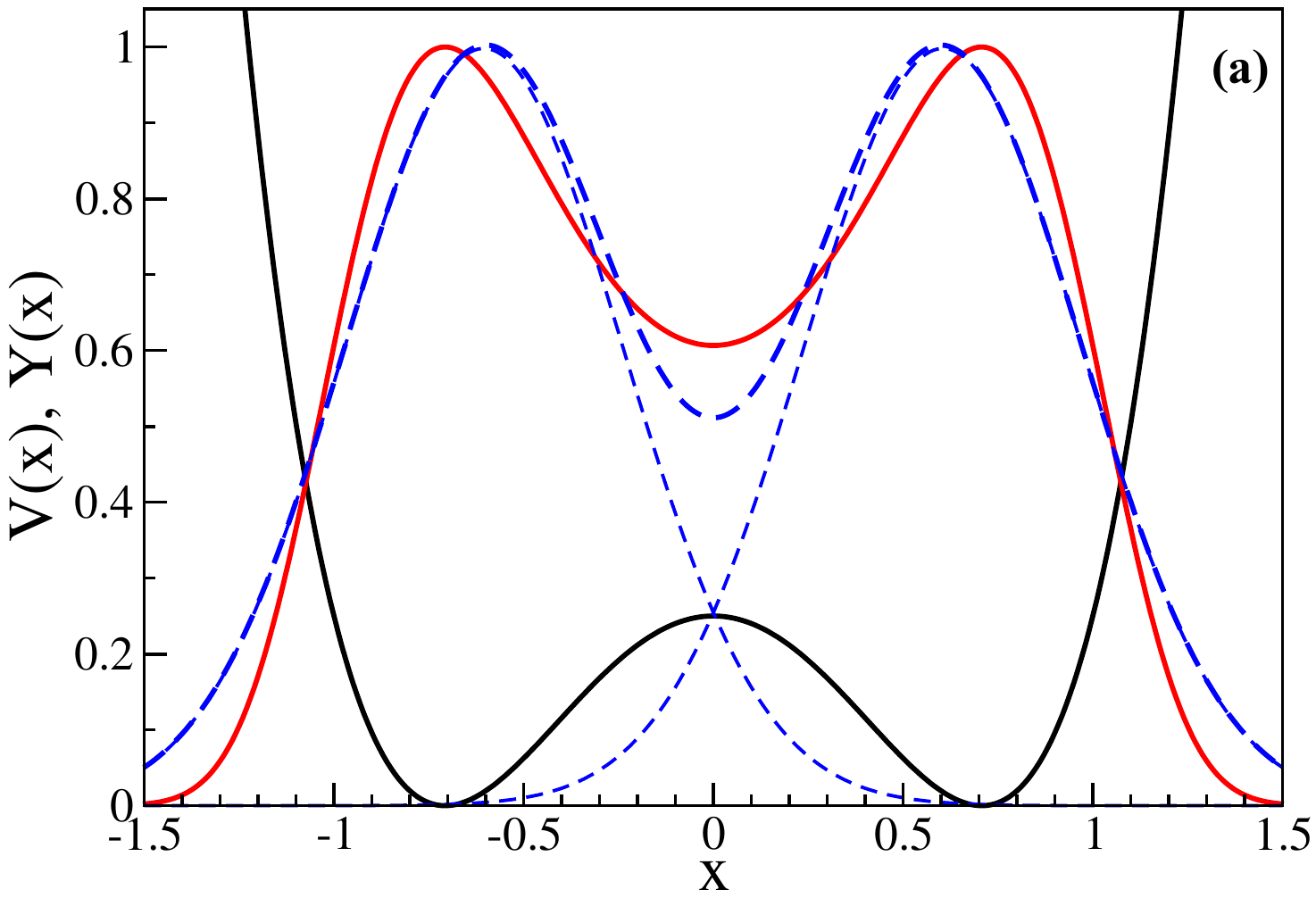}
		\includegraphics[width=0.45\textwidth]{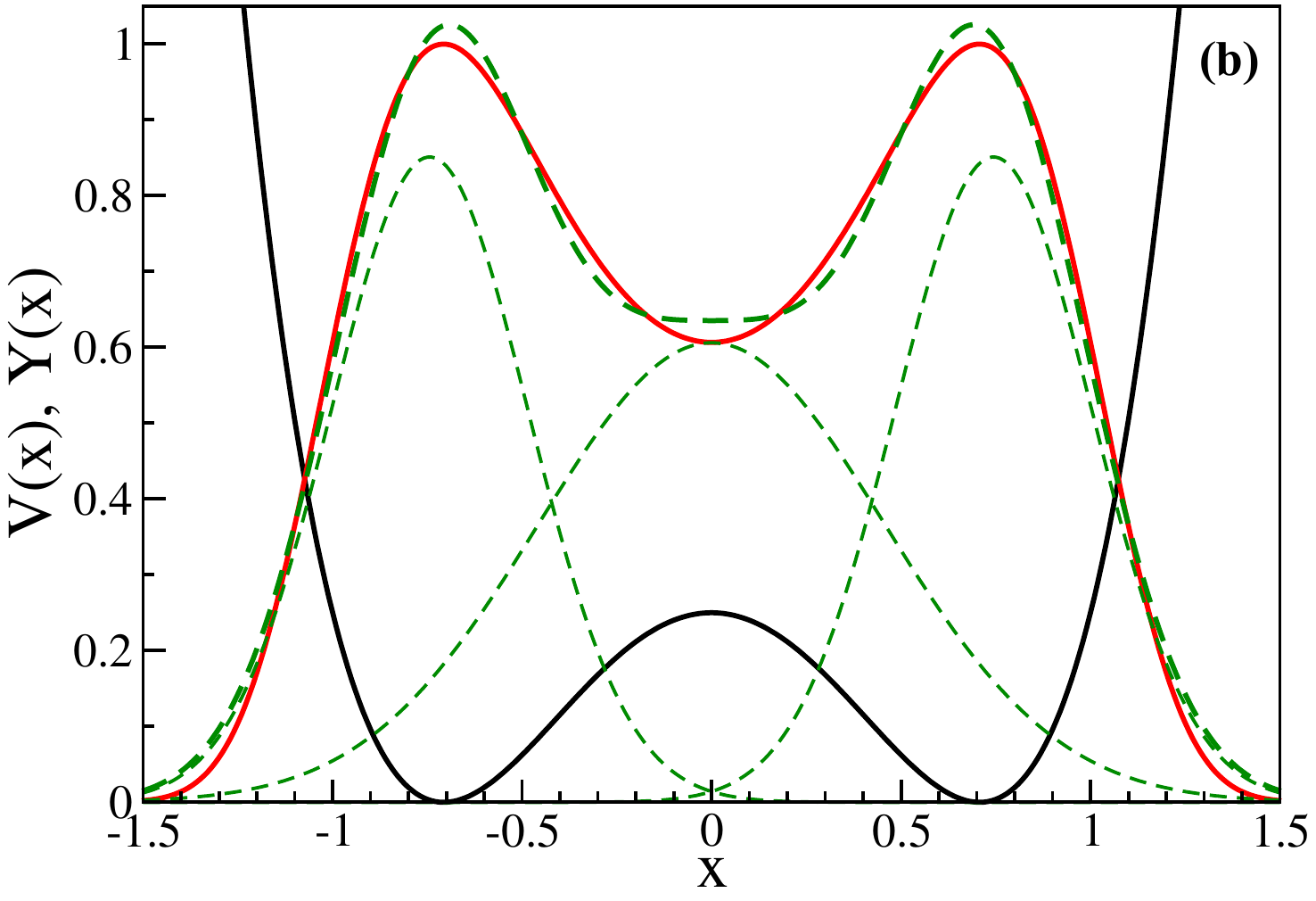}
		\caption{Two-well potential $V=\frac{1}{4}-x^2+x^4$ (black solid line) and associated yield $Y(x)$ from Eq.~(\ref{eq:yield}) with $T=0.5$~MeV  (red solid line). (a) Two- and (b) three-Gaussian fits (thick dashed lines). Thin dashed lines show each individual Gaussian component.}
		\label{fig:2-well}
	\end{figure}

This is illustrated in Fig.~\ref{fig:2-well}(a) and (b) showing the yield at $T=0.5$~MeV together with two- and three-Gaussian fits, respectively. 
The two-Gaussian fits is of poor quality and in particular does not reproduce the peak position well. 
Adding a third  Gaussian at symmetry considerably improves the fit. 
However, the symmetric Gaussian mode should not be misinterpreted as a symmetric theoretical mode. As before, analysis of the covariance and correlation of the fit parameters (Tab.~\ref{tab:Correlation2}) suggest the three-Gaussian fit has too many free parameters to be constrained by these data. 
\begin{table}
    \centering
\begin{tabular}{r|rr}
2-Gaussian&$\sigma_1$ & $x_0$ \\
\hline
$\sigma_1$&1.0529e-04 & 2.8418e-05 \\
$x_0$& \textbf{0.2718}& 1.0385e-04 \\
\end{tabular}

\begin{tabular}{r|rrrr}
3-Gaussian& $\sigma_1$ & $x_0$ & $y$ & $\sigma_2$ \\
\hline
$\sigma_1$&8.4450e-04 & -7.6818e-04 & -5.4315e-03 & -8.9075e-03 \\
$x_0$ & \textbf{-0.9834} &7.2249e-04 & 5.0259e-03 & 8.2222e-03 \\
$y$ & \textbf{-0.9874} & \textbf{0.9878} & 3.5831e-02 & 5.8747e-02 \\
$\sigma_2$&\textbf{-0.9863 }& \textbf{0.9843} & \textbf{0.9986} & 9.6579e-02 \\
\end{tabular}
    \caption{On-diagonal and upper-right: elements of the covariance matrix for the two- and three-Gaussian fits respectively in Fig.~\ref{fig:2-well}. In bold below the diagonal: correlation matrix elements, defined as $C_{ij}=\text{cov}(x_i,x_j)/\sqrt{\text{cov}(x_i,x_i)\text{cov}(x_j,x_j)}$. The covariances are calculated on the assumption of a uniform uncertainty, with data points evaluated at intervals of $0.1$ in $x$ from -1.5 to 1.5.}
    \label{tab:Correlation2}
\end{table}

	\begin{figure}
		\includegraphics[width=0.45\textwidth]{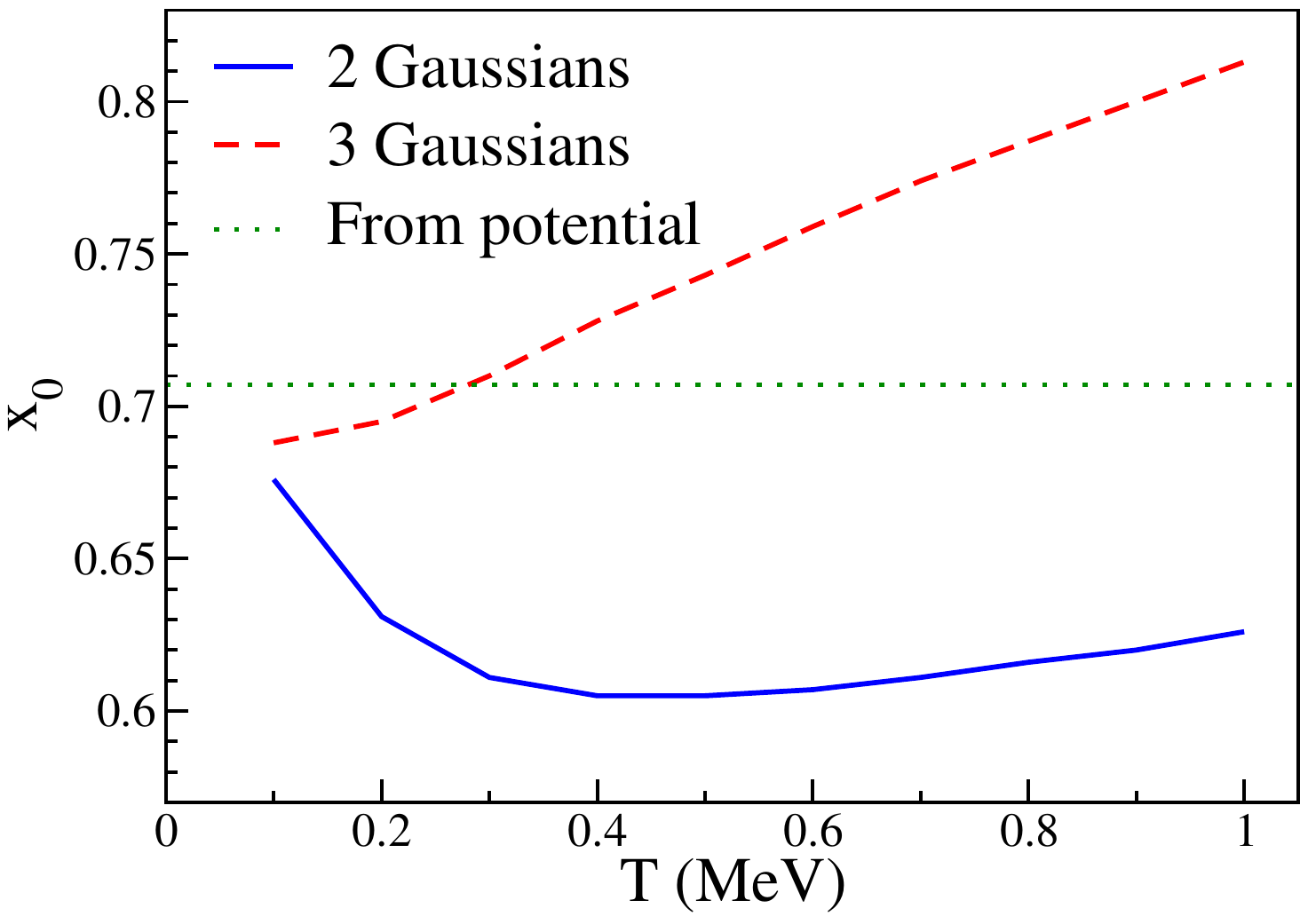}
		\caption{Centroids of the asymmetric Gaussian modes obtained from two- and three-Gaussian fits as a function of $T$ for the two-well potential of Fig.~\ref{fig:asym_T} whose theoretical mode is indicated by the dotted line. }
		\label{fig:asym_x0-T}
	\end{figure}

The excitation energy dependence of the position of the asymmetric Gaussian modes is shown in Fig.~\ref{fig:asym_x0-T}.
Both two- and three-Gaussian fits exhibit significant variations and globally fail to predict the exact asymmetry with precision. 
Again, the excitation energy dependence of the Gaussian modes illustrates their difference with theoretical modes and highlights the importance of comparing effective modes rather than Gaussian ones when comparing data measured at different energies.

	\begin{figure}
		\includegraphics[width=0.45\textwidth]{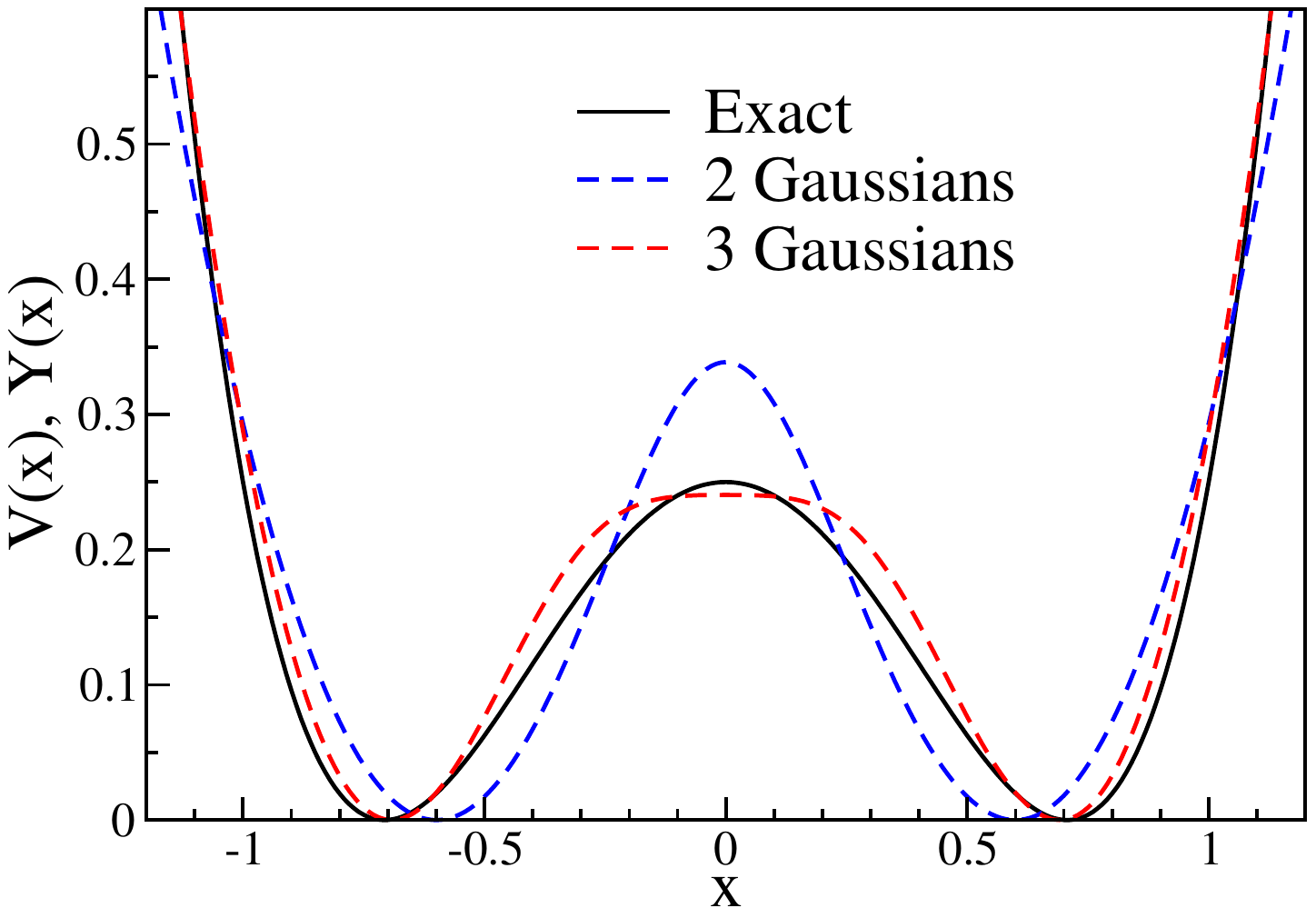}
		\caption{Comparison between the two-well potential (solid line) and the effective potentials extracted from the two (blue dashed line) and three (red dashed line) Gaussian fits in Fig.~\ref{fig:2-well} using Eq.~(\ref{eq:VG}).}
		\label{fig:asym-pot}
	\end{figure}

Figure~\ref{fig:asym-pot} shows the effective potentials extracted from the Gaussian modes using Eq.~(\ref{eq:VG}).
Here, both two- and three-Gaussian fits lead to a unique asymmetric effective mode. 
In particular, the symmetric Gaussian mode in the three-Gaussian fit does not translate into an additional symmetric effective mode. 
Nevertheless, the resulting effective potentials exhibit significant differences with the original two-well potential. 
As in the quartic case, the two-well potential can be exactly reproduced from a fit of the yield as given in Eq.~(\ref{eq:Ypolyn}).

\section{From yield to effective potential \label{sec:YtoV}}

We now illustrate the method of extracting and analysing effective potentials from experimental yields. 
We consider three systems spanning a range of excitation energies: fusion-fission \cite{Nishio2015} and $\beta-$delayed fission \cite{Andreyev2010} of $^{180}$Hg, as well as  neutron induced fission of $^{238}$U and $^{232}$Th \cite{Simutkin2014}. 
Rather than using experimental data as a starting point, we use the Gaussian fits of the fragment mass distributions as obtained by the authors. 
While $^{180}$Hg is well described by one asymmetric Gaussian mode, the actinides require two asymmetric and one symmetric Gaussian modes, also known as Standard 1 (S1), Standard 2 (S2), and symmetric superlong Brosa modes, respectively \cite{Brosa1990}.

\subsection{$^{180}$Hg}

\begin{figure}
		\includegraphics[width=0.45\textwidth]{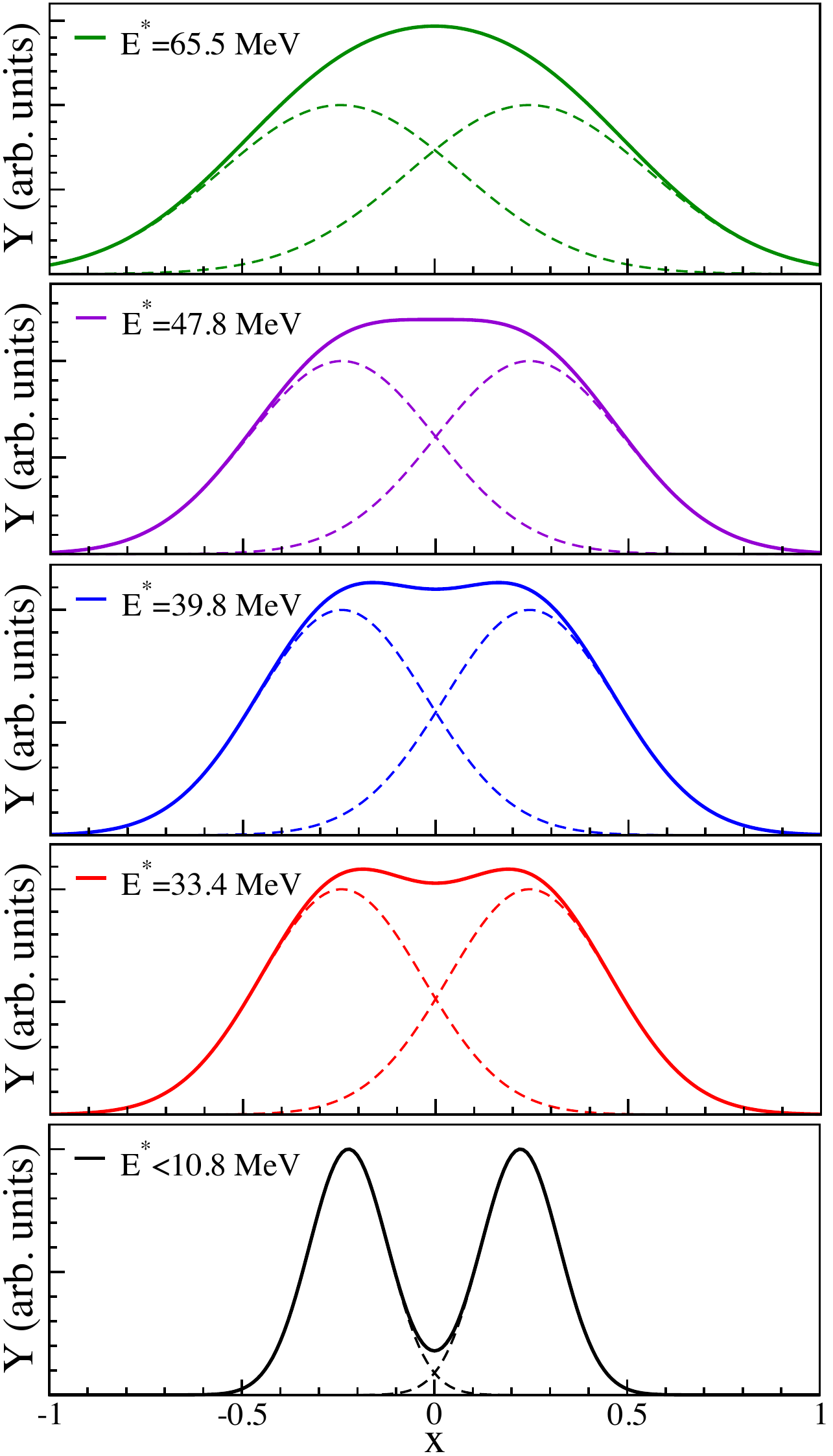}
		\caption{Two Gaussian fits (corresponding to one asymmetric Gaussian mode) of $^{180}$Hg experimental fission yields. The lowest excitation energy is from $\beta-$delayed fission \cite{Andreyev2010} while the others are from $^{36}$Ar$+^{144}$Sm fusion-fission reactions \cite{Nishio2015}.}
		\label{fig:180Hg_yield}
	\end{figure}

 $^{180}$Hg fission fragment mass distributions were reported in \cite{Andreyev2010}  for $\beta-$delayed fission of $^{180}$Tl experiments and in \cite{Nishio2015} for fusion-fission of $^{36}$Ar$+^{144}$Sm  (see also \cite{prasad2015} for similar reactions), spanning a broad range of  excitation energies up to $E^*=65.5$~MeV. These yields were well described by two Gaussian fits \cite{Nishio2015} that are reproduced in Fig.~\ref{fig:180Hg_yield}. The position of the Gaussian centroids was kept constant over the excitation energies \cite{Nishio2015}.
 The asymmetry in the yield distribution, which is attributed to the influence of shell effects, disappears at the highest excitation energy $E^*=65.5$~MeV. 
 Indeed, shell effects are expected to wash out at such high excitation energy. 
 The resulting broad symmetric distribution can be equally well reproduced by a single Gaussian function.
 In this case, the two-Gaussian centroids cannot be anymore interpreted in terms of underlying shell effects.

\begin{figure}
		\includegraphics[width=0.45\textwidth]{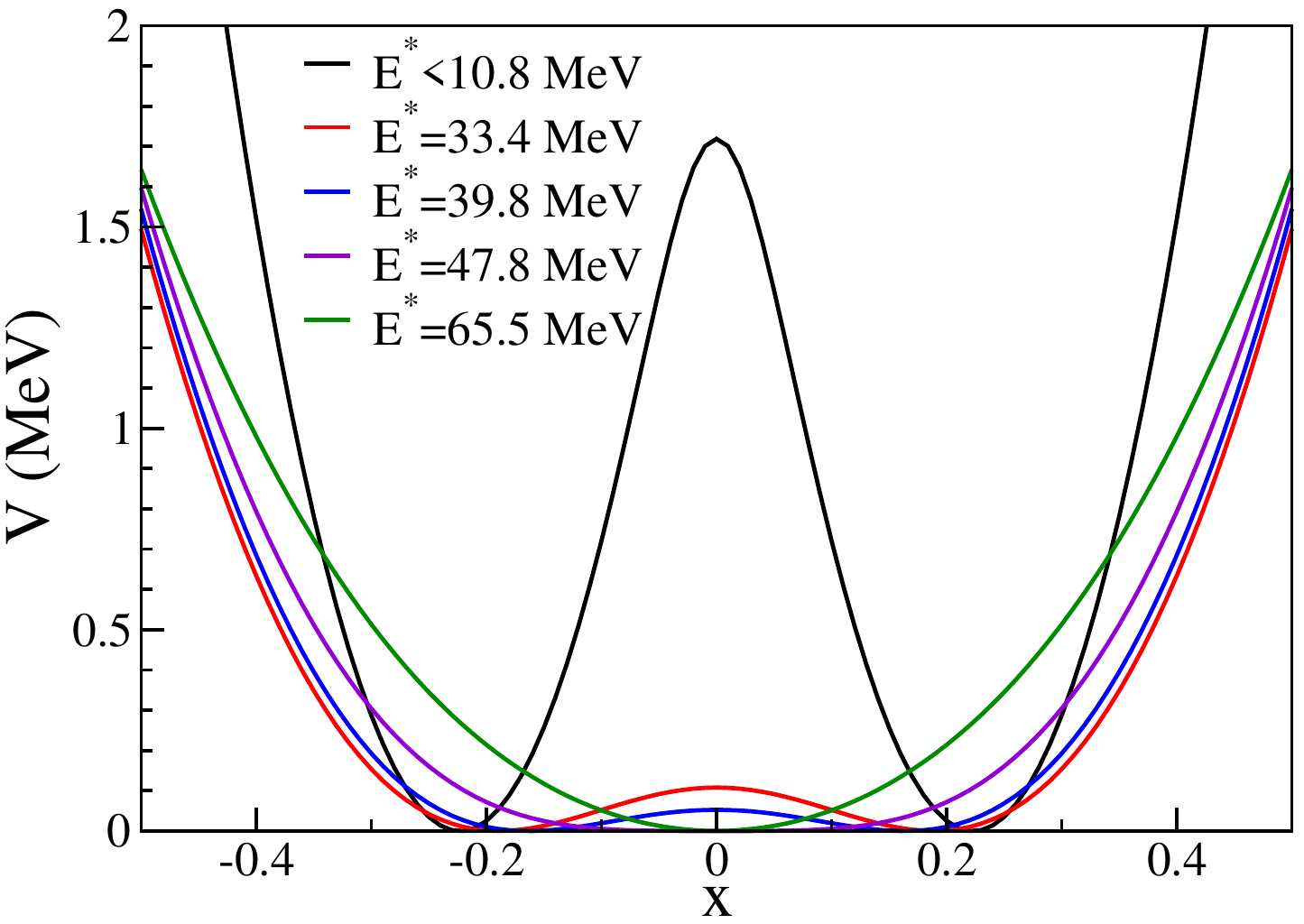}
		\caption{$^{180}$Hg effective potentials from the yields of Fig.~\ref{fig:180Hg_yield}.}
		\label{fig:180Hg_pot}
	\end{figure}

Figure~\ref{fig:180Hg_pot} shows that the effective potentials associated with these yields vary significantly with excitation energy. 
In particular, they take the shape of a two-well potential at low energy, but transition to a single potential well at higher energy. An analysis based on effective potentials therefore reveals that the influence of the shell effect responsible for the asymmetric mode does not appear to persist to energies above $E^*\approx45$~MeV. Instead, the asymmetric effective mode is replaced by a broad symmetric one, suggesting a transition to a more liquid-drop-like behaviour.

\subsection{$^{238}$U neutron induced fission}

		\begin{figure}
		\includegraphics[width=0.45\textwidth]{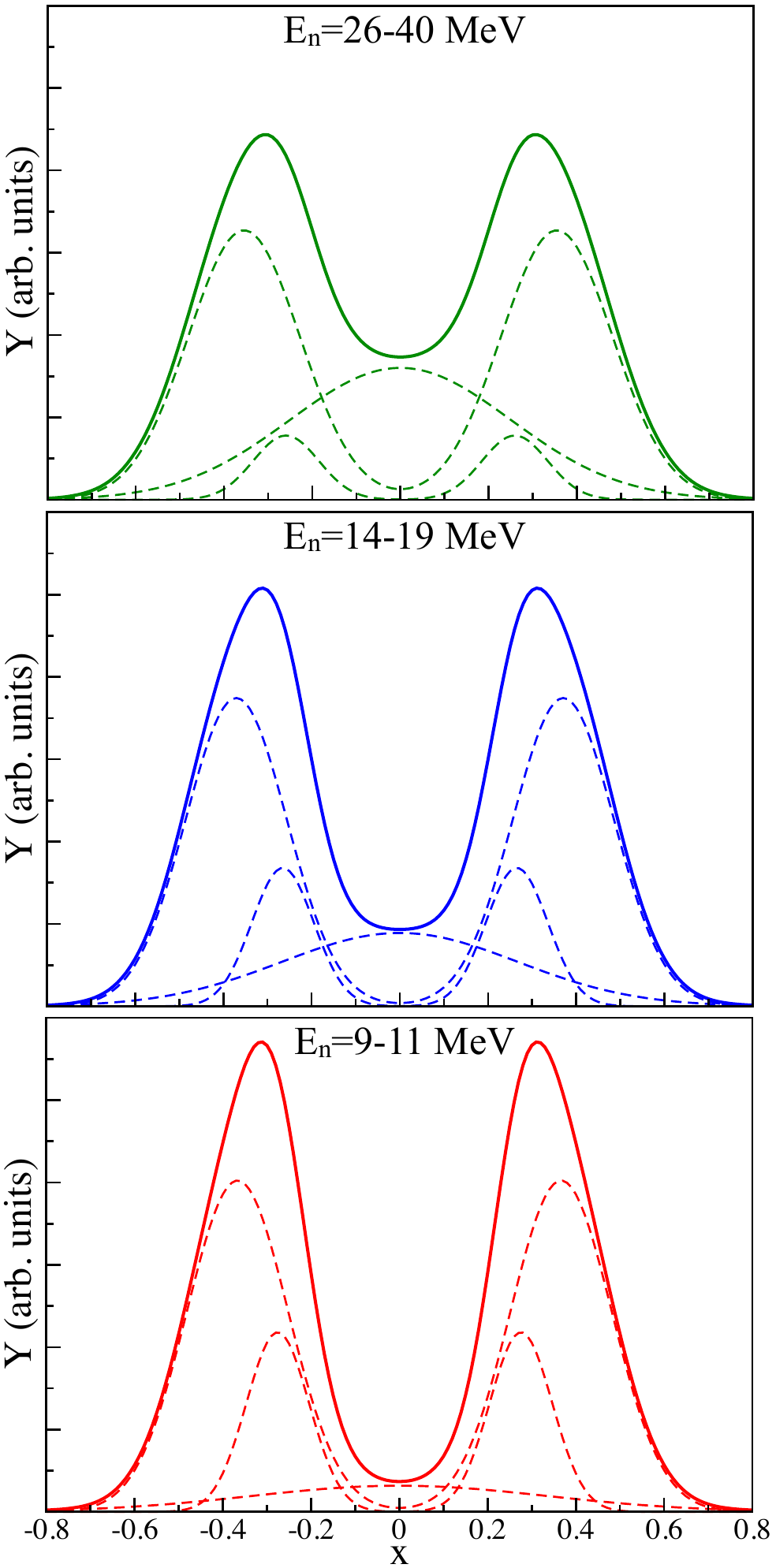}
		\caption{Gaussian modes (dashed lines) and corresponding total yields (solid lines) obtained from fits of experimental fragment mass distributions resulting from $^{238}$U neutron induced fission at various neutron energies \cite{Simutkin2014}.}
		\label{fig:238U_yield}
	\end{figure}

Gaussian modes obtained from $^{238}$U neutron-induced fission data \cite{Simutkin2014} are shown in Fig.~\ref{fig:238U_yield} for various neutron energy ranges. 
As in the $^{180}$Hg case, the centroids of these modes are kept constant across excitation energies. 
The S2  mode, which is the most asymmetric Gaussian mode, clearly dominates the asymmetric peaks.
However, the increase of the yield at symmetry with increasing neutron energy necessitates a symmetric Gaussian mode to achieve a good fit, whose yield  increases with energy.
The latter is very broad and spans a range of $x$ well into the other modes.  

	\begin{figure}
		\includegraphics[width=0.45\textwidth]{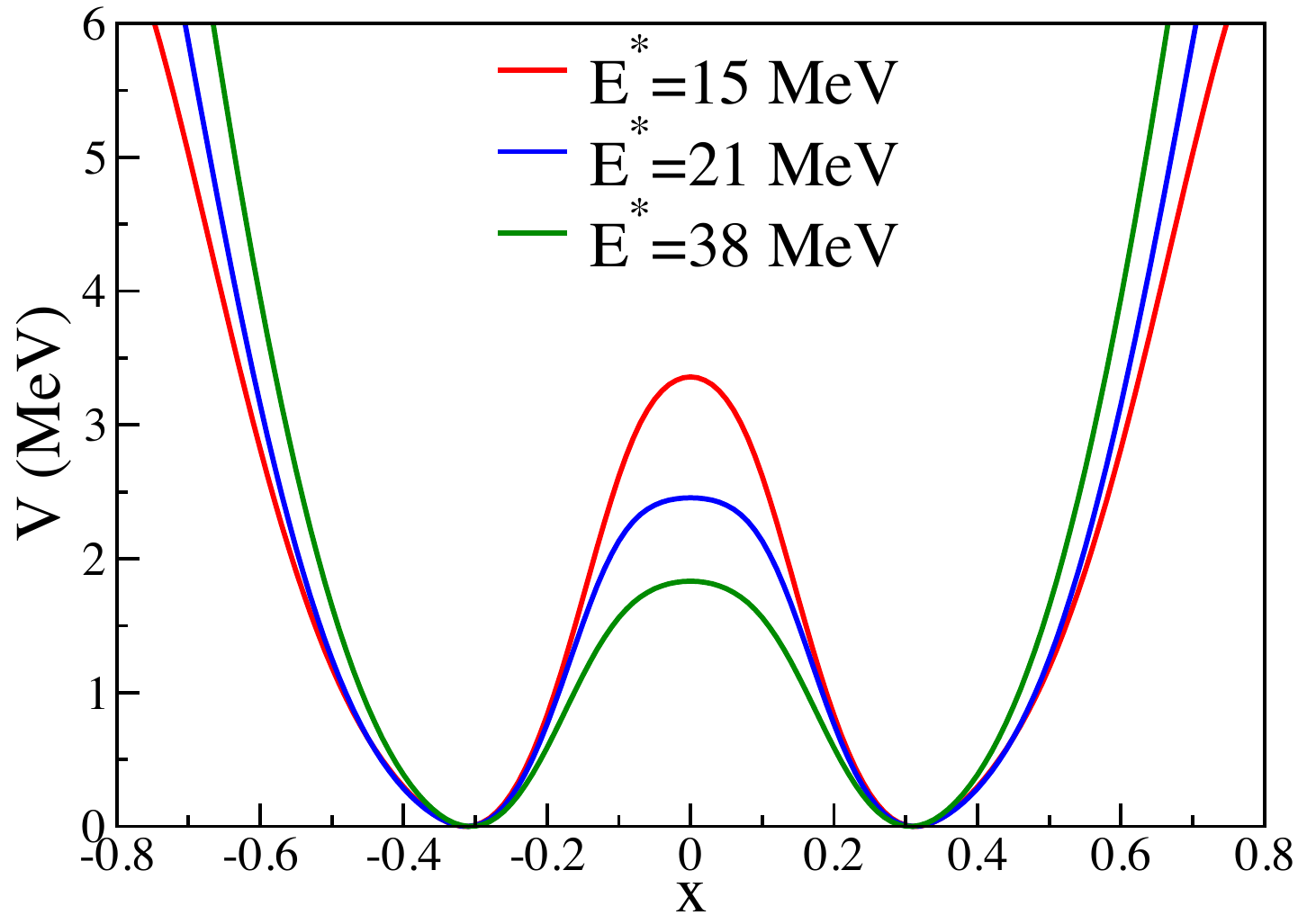}
		\caption{$^{238}$U effective potentials from the total yields of Fig.~\ref{fig:238U_yield}. The average neutron energy was used to evaluate the excitation energies.}
		\label{fig:238U_pot}
	\end{figure}

Figure~\ref{fig:238U_pot} shows the effective potentials at excitation energies corresponding  to the average neutron energy for each range. 
Each effective potential exhibits a clear two-well structure indicating a unique asymmetric effective mode. 
The asymmetry of this mode remains constant with excitation energy, as expected from an influence of shell effects. 
In this interpretation, the increase in yield at symmetry corresponds to a lowering of the potential barrier between the two wells at higher excitation energy.

\subsection{$^{232}$Th neutron induced fission}

		\begin{figure}
		\includegraphics[width=0.45\textwidth]{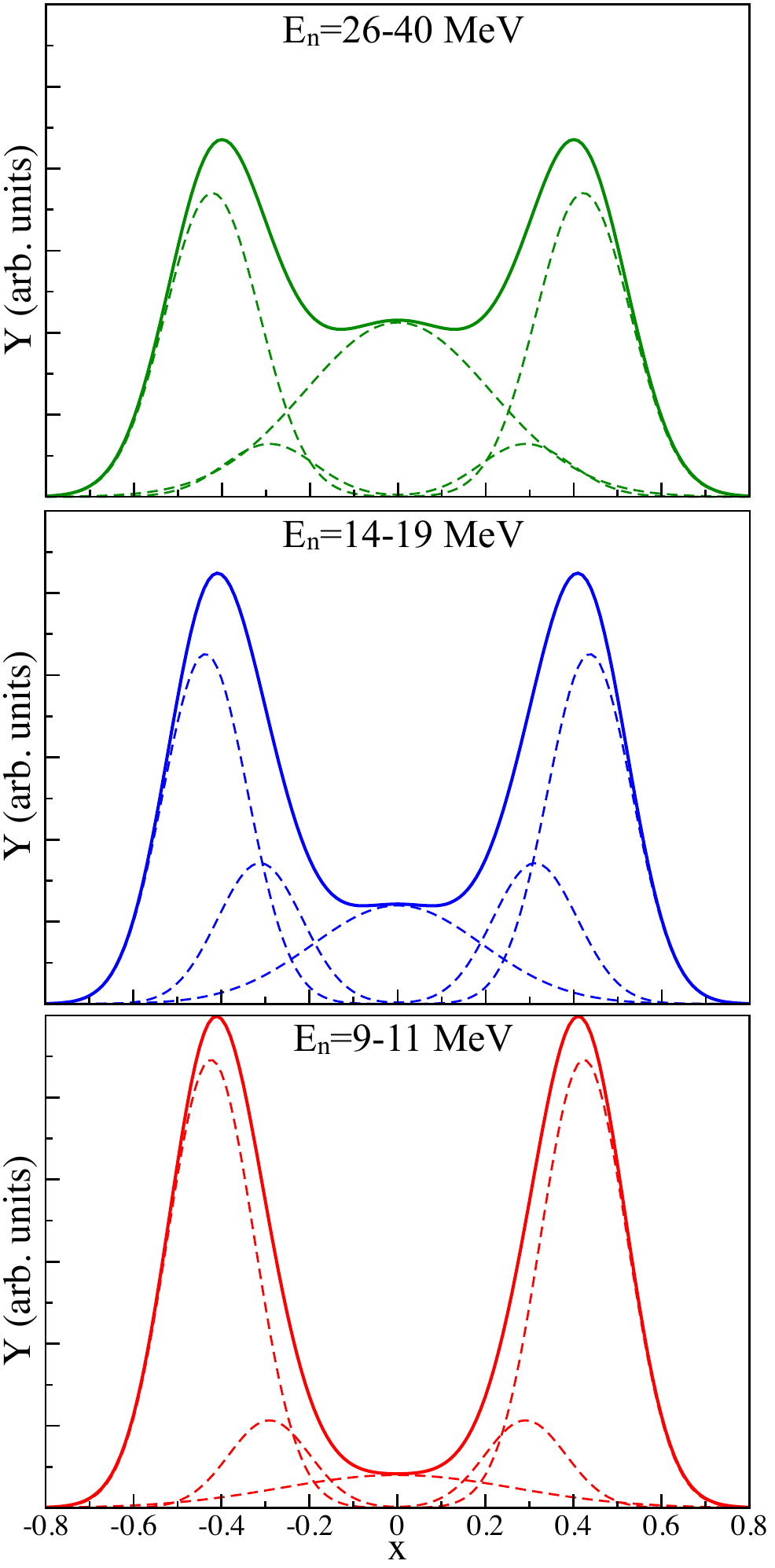}
		\caption{Gaussian modes (dashed lines) and corresponding total yields (solid lines) obtained from fits of experimental fragment mass distributions resulting from $^{232}$Th neutron induced fission at various neutron energies \cite{Simutkin2014}.}
		\label{fig:232Th_yield}
	\end{figure}

\begin{figure}
		\includegraphics[width=0.45\textwidth]{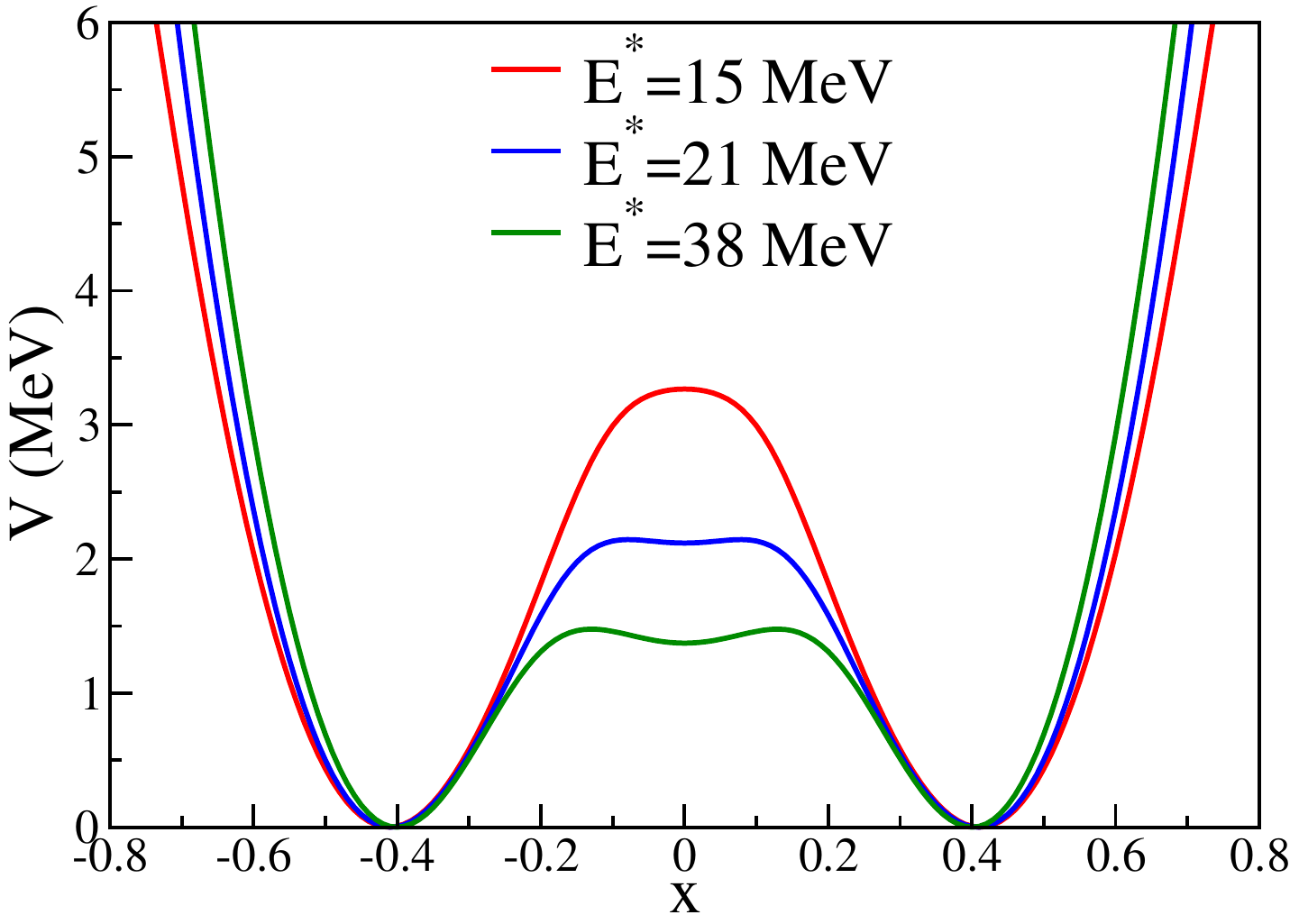}
		\caption{$^{232}$Th effective potentials from the total yields of Fig.~\ref{fig:232Th_yield}. The average neutron energy was used to evaluate the excitation energies.}
		\label{fig:232Th_pot}
	\end{figure}

A similar analysis of Gaussian modes for neutron induced fission on $^{232}$Th at various neutron energies \cite{Simutkin2014} is presented in Fig.~\ref{fig:232Th_yield}. 
 The same S1, S2 and symmetric Gaussian modes are used in the fits. 
 As in the case of $^{238}$U, the asymmetric peak is dominated by the S2 mode. 
 At higher energy, however, the symmetric Gaussian mode becomes significant enough to induce a peak at symmetry.
 The effective potentials shown in Fig.~\ref{fig:232Th_pot} indeed indicate the appearance of a shallow well at symmetry at high $E^*$. 
 
\section{Discussion and conclusion \label{sec:conc}}


In actinides, the S1 and S2 modes are centred around $Z\approx52.5$ and $Z\approx55$ protons in the heavy fragment, respectively \cite{bockstiegel2008}. 
A possible interpretation of the underlying shell effects is that they originate from octupole deformed shell gaps at $Z=52 $ and $Z=56$ \cite{scamps2018}.
Indeed, near scission, the fragments acquire pear shapes induced by the competition between the short range attraction in the neck and the long range repulsion between them. 
The octupole deformed shell effects then facilitate these deformations. 

Despite their different characteristics, the S1 and S2 asymmetric Gaussian modes produce a {\it single} asymmetric effective mode. 
The position of the latter  remains stable with excitation energy, confirming its shell effect origin. 
This observation is in agreement with theoretical PES which, to our knowledge, do not exhibit different valleys for S1 and S2 Gaussian modes (see, e.g., \cite{Goutte2005}), indicating that they both contribute to the same ``Standard'' theoretical mode. 
Naturally, the combination of shell effects responsible for S1 and S2 Gaussian modes are expected to induce a non-harmonic shape of the potential energy at scission.  
Note that the ``supershort" \cite{Brosa1990} Gaussian mode in neutron-rich fermium, attributed to $Z=50$ spherical shell effects, is associated with a different theoretical mode than the standard one, i.e., two valleys separated by a potential ridge are usually found in theoretical PES \cite{bonneau2006,bernard2023}.

The fusion-fission $^{180}$Hg data extend to relatively high excitation energies, for which a transition from an asymmetric to a symmetric effective mode is observed. The role of shell effects on $^{180}$Hg fission fragment formation has been  studied in several theoretical works \cite{Moller2012,ichikawa2012,McDonnell2014,Andreev2012,andreev2013,panebianco2012,warda2012,scamps2019,bernard2024}.
The fact that the effective potential varies with excitation energy is not surprising as the shell effects are expected to weaken with increasing excitation energy,
inducing  the system to transition from a two-well to a one-well effective potential. This behaviour would not be noticed without a calculation of the effective potential, and indeed an analysis based solely on Gaussian modes with fixed centroids would suggest the persistence of this shell effect to at least $E^*=65.5$~MeV.

The $^{238}$U data show an increase of the symmetric Gaussian mode and a decrease of the barrier separating the asymmetric wells, with increasing excitation energy. 
This could also be interpreted in terms of reduction of shell effects. The case of $^{232}$Th, however,  provides indication that  an additional  shallow symmetric well in the effective potential appears at high excitation energy. 
A possible interpretation is that the system now has enough energy to overcome a barrier in the PES leading to a symmetric fission valley. 
The fact that a symmetric well is not observed in $^{238}$U data could be due to a higher saddle point preventing the system to enter the symmetric valley, which is often predicted by theory (see, e.g., \cite{Goutte2005}). 
This is an obvious limitation of using  the effective potential to compare with theoretical predictions when only a limited range of excitation energies is available. 
Although, in principle, the comparison is valid within the framework of the scission point model, it neglects the pre-scission dynamics and non-trivial PES topography. 
The latter require more advanced approaches such as TDGCM \cite{Goutte2005,Schunck2016}.

Summarising, Gaussian fits of fission fragment mass or charge distributions have been a standard tool for decades to identify the role of shell effects in fragment formation. 
However, these ``Gaussian fission modes'' (as, e.g., the S1 and S2 Brosa modes) often contribute to the same potential well at scission (or fission valley in the PES), such that there is not a one-to-one comparison available between ``Gaussian modes" and ``theoretical modes''.
We proposed a simple way to convert fragment yields into an effective potential at scission, that can then be used to define ``effective fission modes''. 
Assuming that a broad range of excitation energies is available, these effective modes provide an easy way to compare with theoretical modes predictions. 

	
\bibliography{sources.bib}

\begin{thebibliography}{71}
\expandafter\ifx\csname natexlab\endcsname\relax\def\natexlab#1{#1}\fi
\expandafter\ifx\csname bibnamefont\endcsname\relax
  \def\bibnamefont#1{#1}\fi
\expandafter\ifx\csname bibfnamefont\endcsname\relax
  \def\bibfnamefont#1{#1}\fi
\expandafter\ifx\csname citenamefont\endcsname\relax
  \def\citenamefont#1{#1}\fi
\expandafter\ifx\csname url\endcsname\relax
  \def\url#1{\texttt{#1}}\fi
\expandafter\ifx\csname urlprefix\endcsname\relax\def\urlprefix{URL }\fi
\providecommand{\bibinfo}[2]{#2}
\providecommand{\eprint}[2][]{\url{#2}}

\bibitem[{\citenamefont{Meitner and Frisch}(1939)}]{Meitner1939}
\bibinfo{author}{\bibfnamefont{L.}~\bibnamefont{Meitner}} \bibnamefont{and} \bibinfo{author}{\bibfnamefont{O.}~\bibnamefont{Frisch}}, \bibinfo{journal}{Nature} \textbf{\bibinfo{volume}{143}}, \bibinfo{pages}{471} (\bibinfo{year}{1939}).

\bibitem[{\citenamefont{Grummitt and Wilkinson}(1946)}]{Grummitt1946}
\bibinfo{author}{\bibfnamefont{W.~E.} \bibnamefont{Grummitt}} \bibnamefont{and} \bibinfo{author}{\bibfnamefont{G.}~\bibnamefont{Wilkinson}}, \bibinfo{journal}{Nature} \textbf{\bibinfo{volume}{158}}, \bibinfo{pages}{163} (\bibinfo{year}{1946}).

\bibitem[{\citenamefont{Turkevich and Niday}(1951)}]{Turkevich1951}
\bibinfo{author}{\bibfnamefont{A.}~\bibnamefont{Turkevich}} \bibnamefont{and} \bibinfo{author}{\bibfnamefont{J.~B.} \bibnamefont{Niday}}, \bibinfo{journal}{Physical Review} \textbf{\bibinfo{volume}{84}}, \bibinfo{pages}{52} (\bibinfo{year}{1951}).

\bibitem[{\citenamefont{Britt et~al.}(1963)\citenamefont{Britt, Wegner, and Gursky}}]{Britt1963}
\bibinfo{author}{\bibfnamefont{H.~C.} \bibnamefont{Britt}}, \bibinfo{author}{\bibfnamefont{H.~E.} \bibnamefont{Wegner}}, \bibnamefont{and} \bibinfo{author}{\bibfnamefont{J.~C.} \bibnamefont{Gursky}}, \bibinfo{journal}{Physical Review} \textbf{\bibinfo{volume}{129}}, \bibinfo{pages}{2239} (\bibinfo{year}{1963}).

\bibitem[{\citenamefont{Gustafsson et~al.}(1971)\citenamefont{Gustafsson, M\"oller, and Nilsson}}]{gustafsson1971}
\bibinfo{author}{\bibfnamefont{C.}~\bibnamefont{Gustafsson}}, \bibinfo{author}{\bibfnamefont{P.}~\bibnamefont{M\"oller}}, \bibnamefont{and} \bibinfo{author}{\bibfnamefont{S.~G.} \bibnamefont{Nilsson}}, \bibinfo{journal}{Phys. Lett. B} \textbf{\bibinfo{volume}{34}}, \bibinfo{pages}{349} (\bibinfo{year}{1971}).

\bibitem[{\citenamefont{Ichikawa and Möller}(2019)}]{ichikawa2019}
\bibinfo{author}{\bibfnamefont{T.}~\bibnamefont{Ichikawa}} \bibnamefont{and} \bibinfo{author}{\bibfnamefont{P.}~\bibnamefont{Möller}}, \bibinfo{journal}{Phys. Lett. B} \textbf{\bibinfo{volume}{789}}, \bibinfo{pages}{679} (\bibinfo{year}{2019}).

\bibitem[{\citenamefont{Ćwiok et~al.}(1994)\citenamefont{Ćwiok, Nazarewicz, Saladin, Płóciennik, and Johnson}}]{cwiok1994}
\bibinfo{author}{\bibfnamefont{S.}~\bibnamefont{Ćwiok}}, \bibinfo{author}{\bibfnamefont{W.}~\bibnamefont{Nazarewicz}}, \bibinfo{author}{\bibfnamefont{J.}~\bibnamefont{Saladin}}, \bibinfo{author}{\bibfnamefont{W.}~\bibnamefont{Płóciennik}}, \bibnamefont{and} \bibinfo{author}{\bibfnamefont{A.}~\bibnamefont{Johnson}}, \bibinfo{journal}{Phys. Lett. B} \textbf{\bibinfo{volume}{322}}, \bibinfo{pages}{304} (\bibinfo{year}{1994}).

\bibitem[{\citenamefont{{Bernard, R. N.} et~al.}(2023)\citenamefont{{Bernard, R. N.}, {Simenel, C.}, and {Blanchon, G.}}}]{bernard2023}
\bibinfo{author}{\bibnamefont{{Bernard, R. N.}}}, \bibinfo{author}{\bibnamefont{{Simenel, C.}}}, \bibnamefont{and} \bibinfo{author}{\bibnamefont{{Blanchon, G.}}}, \bibinfo{journal}{Eur. Phys. J. A} \textbf{\bibinfo{volume}{59}}, \bibinfo{pages}{51} (\bibinfo{year}{2023}), \urlprefix\url{https://doi.org/10.1140/epja/s10050-023-00964-2}.

\bibitem[{\citenamefont{Wilkins et~al.}(1976)\citenamefont{Wilkins, Steinberg, and Chasman}}]{Wilkins1976}
\bibinfo{author}{\bibfnamefont{B.~D.} \bibnamefont{Wilkins}}, \bibinfo{author}{\bibfnamefont{E.~P.} \bibnamefont{Steinberg}}, \bibnamefont{and} \bibinfo{author}{\bibfnamefont{R.~R.} \bibnamefont{Chasman}}, \bibinfo{journal}{Phys. Rev. C} \textbf{\bibinfo{volume}{14}}, \bibinfo{pages}{1832} (\bibinfo{year}{1976}).

\bibitem[{\citenamefont{Sadhukhan et~al.}(2016)\citenamefont{Sadhukhan, Nazarewicz, and Schunck}}]{Sadhukhan2016}
\bibinfo{author}{\bibfnamefont{J.}~\bibnamefont{Sadhukhan}}, \bibinfo{author}{\bibfnamefont{W.}~\bibnamefont{Nazarewicz}}, \bibnamefont{and} \bibinfo{author}{\bibfnamefont{N.}~\bibnamefont{Schunck}}, \bibinfo{journal}{Phys. Rev. C} \textbf{\bibinfo{volume}{93}}, \bibinfo{pages}{011304} (\bibinfo{year}{2016}).

\bibitem[{\citenamefont{Scamps and Simenel}(2018)}]{scamps2018}
\bibinfo{author}{\bibfnamefont{G.}~\bibnamefont{Scamps}} \bibnamefont{and} \bibinfo{author}{\bibfnamefont{C.}~\bibnamefont{Simenel}}, \bibinfo{journal}{Nature} \textbf{\bibinfo{volume}{564}}, \bibinfo{pages}{382} (\bibinfo{year}{2018}).

\bibitem[{\citenamefont{Scamps and Simenel}(2019)}]{scamps2019}
\bibinfo{author}{\bibfnamefont{G.}~\bibnamefont{Scamps}} \bibnamefont{and} \bibinfo{author}{\bibfnamefont{C.}~\bibnamefont{Simenel}}, \bibinfo{journal}{Phys. Rev. C} \textbf{\bibinfo{volume}{100}}, \bibinfo{pages}{041602(R)} (\bibinfo{year}{2019}).

\bibitem[{\citenamefont{Mahata et~al.}(2022)\citenamefont{Mahata, Schmitt, Gupta, Shrivastava, Scamps, and Schmidt}}]{mahata2022}
\bibinfo{author}{\bibfnamefont{K.}~\bibnamefont{Mahata}}, \bibinfo{author}{\bibfnamefont{C.}~\bibnamefont{Schmitt}}, \bibinfo{author}{\bibfnamefont{S.}~\bibnamefont{Gupta}}, \bibinfo{author}{\bibfnamefont{A.}~\bibnamefont{Shrivastava}}, \bibinfo{author}{\bibfnamefont{G.}~\bibnamefont{Scamps}}, \bibnamefont{and} \bibinfo{author}{\bibfnamefont{K.-H.} \bibnamefont{Schmidt}}, \bibinfo{journal}{Phys. Lett. B} \textbf{\bibinfo{volume}{825}}, \bibinfo{pages}{136859} (\bibinfo{year}{2022}).

\bibitem[{\citenamefont{Brosa et~al.}(1990)\citenamefont{Brosa, Grossmann, and M{\"u}ller}}]{Brosa1990}
\bibinfo{author}{\bibfnamefont{U.}~\bibnamefont{Brosa}}, \bibinfo{author}{\bibfnamefont{S.}~\bibnamefont{Grossmann}}, \bibnamefont{and} \bibinfo{author}{\bibfnamefont{A.}~\bibnamefont{M{\"u}ller}}, \bibinfo{journal}{Physics Reports} \textbf{\bibinfo{volume}{197}}, \bibinfo{pages}{167} (\bibinfo{year}{1990}).

\bibitem[{\citenamefont{Caama{\~n}o et~al.}(2015)\citenamefont{Caama{\~n}o, Farget, Delaune, Schmidt, Schmitt, Audouin, Bacri, Benlliure, Casarejos, Derkx et~al.}}]{Caamano2015}
\bibinfo{author}{\bibfnamefont{M.}~\bibnamefont{Caama{\~n}o}}, \bibinfo{author}{\bibfnamefont{F.}~\bibnamefont{Farget}}, \bibinfo{author}{\bibfnamefont{O.}~\bibnamefont{Delaune}}, \bibinfo{author}{\bibfnamefont{K.-H.} \bibnamefont{Schmidt}}, \bibinfo{author}{\bibfnamefont{C.}~\bibnamefont{Schmitt}}, \bibinfo{author}{\bibfnamefont{L.}~\bibnamefont{Audouin}}, \bibinfo{author}{\bibfnamefont{C.-O.} \bibnamefont{Bacri}}, \bibinfo{author}{\bibfnamefont{J.}~\bibnamefont{Benlliure}}, \bibinfo{author}{\bibfnamefont{E.}~\bibnamefont{Casarejos}}, \bibinfo{author}{\bibfnamefont{X.}~\bibnamefont{Derkx}}, \bibnamefont{et~al.}, \bibinfo{journal}{Physical Review C} \textbf{\bibinfo{volume}{92}}, \bibinfo{pages}{034606} (\bibinfo{year}{2015}).

\bibitem[{\citenamefont{Caama{\~n}o et~al.}(2013)\citenamefont{Caama{\~n}o, Delaune, Farget, Derkx, Schmidt, Audouin, Bacri, Barreau, Benlliure, Casarejos et~al.}}]{Caamano2013}
\bibinfo{author}{\bibfnamefont{M.}~\bibnamefont{Caama{\~n}o}}, \bibinfo{author}{\bibfnamefont{O.}~\bibnamefont{Delaune}}, \bibinfo{author}{\bibfnamefont{F.}~\bibnamefont{Farget}}, \bibinfo{author}{\bibfnamefont{X.}~\bibnamefont{Derkx}}, \bibinfo{author}{\bibfnamefont{K.-H.} \bibnamefont{Schmidt}}, \bibinfo{author}{\bibfnamefont{L.}~\bibnamefont{Audouin}}, \bibinfo{author}{\bibfnamefont{C.-O.} \bibnamefont{Bacri}}, \bibinfo{author}{\bibfnamefont{G.}~\bibnamefont{Barreau}}, \bibinfo{author}{\bibfnamefont{J.}~\bibnamefont{Benlliure}}, \bibinfo{author}{\bibfnamefont{E.}~\bibnamefont{Casarejos}}, \bibnamefont{et~al.}, \bibinfo{journal}{Physical Review C} \textbf{\bibinfo{volume}{88}}, \bibinfo{pages}{024605} (\bibinfo{year}{2013}).

\bibitem[{\citenamefont{Schmidt et~al.}(2000)\citenamefont{Schmidt, Steinh{\"a}user, B{\"o}ckstiegel, Grewe, Heinz, Junghans, Benlliure, Clerc, {de Jong}, M{\"u}ller et~al.}}]{Schmidt2000}
\bibinfo{author}{\bibfnamefont{K.-H.} \bibnamefont{Schmidt}}, \bibinfo{author}{\bibfnamefont{S.}~\bibnamefont{Steinh{\"a}user}}, \bibinfo{author}{\bibfnamefont{C.}~\bibnamefont{B{\"o}ckstiegel}}, \bibinfo{author}{\bibfnamefont{A.}~\bibnamefont{Grewe}}, \bibinfo{author}{\bibfnamefont{A.}~\bibnamefont{Heinz}}, \bibinfo{author}{\bibfnamefont{A.}~\bibnamefont{Junghans}}, \bibinfo{author}{\bibfnamefont{J.}~\bibnamefont{Benlliure}}, \bibinfo{author}{\bibfnamefont{H.-G.} \bibnamefont{Clerc}}, \bibinfo{author}{\bibfnamefont{M.}~\bibnamefont{{de Jong}}}, \bibinfo{author}{\bibfnamefont{J.}~\bibnamefont{M{\"u}ller}}, \bibnamefont{et~al.}, \bibinfo{journal}{Nuclear Physics A} \textbf{\bibinfo{volume}{665}}, \bibinfo{pages}{221} (\bibinfo{year}{2000}).

\bibitem[{\citenamefont{Schmidt et~al.}(2001)\citenamefont{Schmidt, Benlliure, and Junghans}}]{Schmidt2001}
\bibinfo{author}{\bibfnamefont{K.~H.} \bibnamefont{Schmidt}}, \bibinfo{author}{\bibfnamefont{J.}~\bibnamefont{Benlliure}}, \bibnamefont{and} \bibinfo{author}{\bibfnamefont{A.~R.} \bibnamefont{Junghans}}, \bibinfo{journal}{Nuclear Physics A} \textbf{\bibinfo{volume}{693}}, \bibinfo{pages}{169} (\bibinfo{year}{2001}).

\bibitem[{\citenamefont{Fr{\'e}geau et~al.}(2016)\citenamefont{Fr{\'e}geau, Oberstedt, Gamboni, Geerts, Hambsch, and Vidali}}]{Fregeau2016}
\bibinfo{author}{\bibfnamefont{M.~O.} \bibnamefont{Fr{\'e}geau}}, \bibinfo{author}{\bibfnamefont{S.}~\bibnamefont{Oberstedt}}, \bibinfo{author}{\bibfnamefont{{\relax Th}.}~\bibnamefont{Gamboni}}, \bibinfo{author}{\bibfnamefont{W.}~\bibnamefont{Geerts}}, \bibinfo{author}{\bibfnamefont{F.~J.} \bibnamefont{Hambsch}}, \bibnamefont{and} \bibinfo{author}{\bibfnamefont{M.}~\bibnamefont{Vidali}}, \bibinfo{journal}{Nuclear Instruments and Methods in Physics Research Section A: Accelerators, Spectrometers, Detectors and Associated Equipment} \textbf{\bibinfo{volume}{817}}, \bibinfo{pages}{35} (\bibinfo{year}{2016}).

\bibitem[{\citenamefont{Chatillon et~al.}(2019)\citenamefont{Chatillon, Ta{\"i}eb, {Alvarez-Pol}, Audouin, Ayyad, B{\'e}lier, Benlliure, Boutoux, Caama{\~n}o, Casarejos et~al.}}]{Chatillon2019}
\bibinfo{author}{\bibfnamefont{A.}~\bibnamefont{Chatillon}}, \bibinfo{author}{\bibfnamefont{J.}~\bibnamefont{Ta{\"i}eb}}, \bibinfo{author}{\bibfnamefont{H.}~\bibnamefont{{Alvarez-Pol}}}, \bibinfo{author}{\bibfnamefont{L.}~\bibnamefont{Audouin}}, \bibinfo{author}{\bibfnamefont{Y.}~\bibnamefont{Ayyad}}, \bibinfo{author}{\bibfnamefont{G.}~\bibnamefont{B{\'e}lier}}, \bibinfo{author}{\bibfnamefont{J.}~\bibnamefont{Benlliure}}, \bibinfo{author}{\bibfnamefont{G.}~\bibnamefont{Boutoux}}, \bibinfo{author}{\bibfnamefont{M.}~\bibnamefont{Caama{\~n}o}}, \bibinfo{author}{\bibfnamefont{E.}~\bibnamefont{Casarejos}}, \bibnamefont{et~al.}, \bibinfo{journal}{Physical Review C} \textbf{\bibinfo{volume}{99}} (\bibinfo{year}{2019}).

\bibitem[{\citenamefont{Nishio et~al.}(2008)\citenamefont{Nishio, Ikezoe, Mitsuoka, Nishinaka, Nagame, Watanabe, Ohtsuki, Hirose, and Hofmann}}]{Nishio2008}
\bibinfo{author}{\bibfnamefont{K.}~\bibnamefont{Nishio}}, \bibinfo{author}{\bibfnamefont{H.}~\bibnamefont{Ikezoe}}, \bibinfo{author}{\bibfnamefont{S.}~\bibnamefont{Mitsuoka}}, \bibinfo{author}{\bibfnamefont{I.}~\bibnamefont{Nishinaka}}, \bibinfo{author}{\bibfnamefont{Y.}~\bibnamefont{Nagame}}, \bibinfo{author}{\bibfnamefont{Y.}~\bibnamefont{Watanabe}}, \bibinfo{author}{\bibfnamefont{T.}~\bibnamefont{Ohtsuki}}, \bibinfo{author}{\bibfnamefont{K.}~\bibnamefont{Hirose}}, \bibnamefont{and} \bibinfo{author}{\bibfnamefont{S.}~\bibnamefont{Hofmann}}, \bibinfo{journal}{Physical Review C} \textbf{\bibinfo{volume}{77}}, \bibinfo{pages}{064607} (\bibinfo{year}{2008}).

\bibitem[{\citenamefont{Chatillon et~al.}(2023)\citenamefont{Chatillon, Boutoux, Gorbinet, Grente, Martin, Pellereau, Taieb, {Alvarez-Pol}, Audouin, Ayyad et~al.}}]{Chatillon2023}
\bibinfo{author}{\bibfnamefont{A.}~\bibnamefont{Chatillon}}, \bibinfo{author}{\bibfnamefont{G.}~\bibnamefont{Boutoux}}, \bibinfo{author}{\bibfnamefont{T.}~\bibnamefont{Gorbinet}}, \bibinfo{author}{\bibfnamefont{L.}~\bibnamefont{Grente}}, \bibinfo{author}{\bibfnamefont{J.-F.} \bibnamefont{Martin}}, \bibinfo{author}{\bibfnamefont{E.}~\bibnamefont{Pellereau}}, \bibinfo{author}{\bibfnamefont{J.}~\bibnamefont{Taieb}}, \bibinfo{author}{\bibfnamefont{H.}~\bibnamefont{{Alvarez-Pol}}}, \bibinfo{author}{\bibfnamefont{L.}~\bibnamefont{Audouin}}, \bibinfo{author}{\bibfnamefont{Y.}~\bibnamefont{Ayyad}}, \bibnamefont{et~al.}, \bibinfo{journal}{EPJ Web of Conferences} \textbf{\bibinfo{volume}{284}}, \bibinfo{pages}{04002} (\bibinfo{year}{2023}).

\bibitem[{\citenamefont{Jhingan et~al.}(2022)\citenamefont{Jhingan, Schmitt, Lemasson, Biswas, Kim, Ramos, Andreyev, Curien, Ciema{\l}a, Cl{\'e}ment et~al.}}]{Jhingan2022}
\bibinfo{author}{\bibfnamefont{A.}~\bibnamefont{Jhingan}}, \bibinfo{author}{\bibfnamefont{C.}~\bibnamefont{Schmitt}}, \bibinfo{author}{\bibfnamefont{A.}~\bibnamefont{Lemasson}}, \bibinfo{author}{\bibfnamefont{S.}~\bibnamefont{Biswas}}, \bibinfo{author}{\bibfnamefont{Y.~H.} \bibnamefont{Kim}}, \bibinfo{author}{\bibfnamefont{D.}~\bibnamefont{Ramos}}, \bibinfo{author}{\bibfnamefont{A.~N.} \bibnamefont{Andreyev}}, \bibinfo{author}{\bibfnamefont{D.}~\bibnamefont{Curien}}, \bibinfo{author}{\bibfnamefont{M.}~\bibnamefont{Ciema{\l}a}}, \bibinfo{author}{\bibfnamefont{E.}~\bibnamefont{Cl{\'e}ment}}, \bibnamefont{et~al.}, \bibinfo{journal}{Physical Review C} \textbf{\bibinfo{volume}{106}}, \bibinfo{pages}{044607} (\bibinfo{year}{2022}).

\bibitem[{\citenamefont{Fern{\'a}ndez et~al.}(2023)\citenamefont{Fern{\'a}ndez, Caama{\~n}o, Ramos, Lemasson, Rejmund, {\'A}lvarez-Pol, Audouin, Frankland, {Fern{\'a}ndez-Dom{\'i}nguez}, {Galiana-Bald{\'o}} et~al.}}]{Fernandez2023}
\bibinfo{author}{\bibfnamefont{D.}~\bibnamefont{Fern{\'a}ndez}}, \bibinfo{author}{\bibfnamefont{M.}~\bibnamefont{Caama{\~n}o}}, \bibinfo{author}{\bibfnamefont{D.}~\bibnamefont{Ramos}}, \bibinfo{author}{\bibfnamefont{A.}~\bibnamefont{Lemasson}}, \bibinfo{author}{\bibfnamefont{M.}~\bibnamefont{Rejmund}}, \bibinfo{author}{\bibfnamefont{H.}~\bibnamefont{{\'A}lvarez-Pol}}, \bibinfo{author}{\bibfnamefont{L.}~\bibnamefont{Audouin}}, \bibinfo{author}{\bibfnamefont{J.~D.} \bibnamefont{Frankland}}, \bibinfo{author}{\bibfnamefont{B.}~\bibnamefont{{Fern{\'a}ndez-Dom{\'i}nguez}}}, \bibinfo{author}{\bibfnamefont{E.}~\bibnamefont{{Galiana-Bald{\'o}}}}, \bibnamefont{et~al.}, \bibinfo{journal}{EPJ Web of Conferences} \textbf{\bibinfo{volume}{290}}, \bibinfo{pages}{02012} (\bibinfo{year}{2023}).

\bibitem[{\citenamefont{Simutkin et~al.}(2014)\citenamefont{Simutkin, Pomp, Blomgren, {\"O}sterlund, Bevilacqua, Andersson, Ryzhov, Tutin, Yavshits, Vaishnene et~al.}}]{Simutkin2014}
\bibinfo{author}{\bibfnamefont{V.~D.} \bibnamefont{Simutkin}}, \bibinfo{author}{\bibfnamefont{S.}~\bibnamefont{Pomp}}, \bibinfo{author}{\bibfnamefont{J.}~\bibnamefont{Blomgren}}, \bibinfo{author}{\bibfnamefont{M.}~\bibnamefont{{\"O}sterlund}}, \bibinfo{author}{\bibfnamefont{R.}~\bibnamefont{Bevilacqua}}, \bibinfo{author}{\bibfnamefont{P.}~\bibnamefont{Andersson}}, \bibinfo{author}{\bibfnamefont{I.~V.} \bibnamefont{Ryzhov}}, \bibinfo{author}{\bibfnamefont{G.~A.} \bibnamefont{Tutin}}, \bibinfo{author}{\bibfnamefont{S.~G.} \bibnamefont{Yavshits}}, \bibinfo{author}{\bibfnamefont{L.~A.} \bibnamefont{Vaishnene}}, \bibnamefont{et~al.}, \bibinfo{journal}{Nuclear Data Sheets} \textbf{\bibinfo{volume}{119}}, \bibinfo{pages}{331} (\bibinfo{year}{2014}).

\bibitem[{\citenamefont{Kozulin et~al.}(2022)\citenamefont{Kozulin, Knyazheva, Itkis, Itkis, Mukhamejanov, Bogachev, Novikov, Kirakosyan, Kumar, Banerjee et~al.}}]{Kozulin2022}
\bibinfo{author}{\bibfnamefont{E.~M.} \bibnamefont{Kozulin}}, \bibinfo{author}{\bibfnamefont{G.~N.} \bibnamefont{Knyazheva}}, \bibinfo{author}{\bibfnamefont{I.~M.} \bibnamefont{Itkis}}, \bibinfo{author}{\bibfnamefont{M.~G.} \bibnamefont{Itkis}}, \bibinfo{author}{\bibfnamefont{Y.~S.} \bibnamefont{Mukhamejanov}}, \bibinfo{author}{\bibfnamefont{A.~A.} \bibnamefont{Bogachev}}, \bibinfo{author}{\bibfnamefont{K.~V.} \bibnamefont{Novikov}}, \bibinfo{author}{\bibfnamefont{V.~V.} \bibnamefont{Kirakosyan}}, \bibinfo{author}{\bibfnamefont{D.}~\bibnamefont{Kumar}}, \bibinfo{author}{\bibfnamefont{T.}~\bibnamefont{Banerjee}}, \bibnamefont{et~al.}, \bibinfo{journal}{Physical Review C} \textbf{\bibinfo{volume}{105}}, \bibinfo{pages}{014607} (\bibinfo{year}{2022}).

\bibitem[{\citenamefont{L{\'e}guillon et~al.}(2016)\citenamefont{L{\'e}guillon, Nishio, Hirose, Makii, Nishinaka, Orlandi, Tsukada, Smallcombe, Chiba, Aritomo et~al.}}]{Leguillon2016a}
\bibinfo{author}{\bibfnamefont{R.}~\bibnamefont{L{\'e}guillon}}, \bibinfo{author}{\bibfnamefont{K.}~\bibnamefont{Nishio}}, \bibinfo{author}{\bibfnamefont{K.}~\bibnamefont{Hirose}}, \bibinfo{author}{\bibfnamefont{H.}~\bibnamefont{Makii}}, \bibinfo{author}{\bibfnamefont{I.}~\bibnamefont{Nishinaka}}, \bibinfo{author}{\bibfnamefont{R.}~\bibnamefont{Orlandi}}, \bibinfo{author}{\bibfnamefont{K.}~\bibnamefont{Tsukada}}, \bibinfo{author}{\bibfnamefont{J.}~\bibnamefont{Smallcombe}}, \bibinfo{author}{\bibfnamefont{S.}~\bibnamefont{Chiba}}, \bibinfo{author}{\bibfnamefont{Y.}~\bibnamefont{Aritomo}}, \bibnamefont{et~al.}, \bibinfo{journal}{Physics Letters B} \textbf{\bibinfo{volume}{761}}, \bibinfo{pages}{125} (\bibinfo{year}{2016}).

\bibitem[{\citenamefont{{Swinton-Bland} et~al.}(2020)\citenamefont{{Swinton-Bland}, Stoyer, Berriman, Hinde, Simenel, Buete, Tanaka, Banerjee, Bezzina, Carter et~al.}}]{Swinton-Bland2020b}
\bibinfo{author}{\bibfnamefont{B.~M.~A.} \bibnamefont{{Swinton-Bland}}}, \bibinfo{author}{\bibfnamefont{M.~A.} \bibnamefont{Stoyer}}, \bibinfo{author}{\bibfnamefont{A.~C.} \bibnamefont{Berriman}}, \bibinfo{author}{\bibfnamefont{D.~J.} \bibnamefont{Hinde}}, \bibinfo{author}{\bibfnamefont{C.}~\bibnamefont{Simenel}}, \bibinfo{author}{\bibfnamefont{J.}~\bibnamefont{Buete}}, \bibinfo{author}{\bibfnamefont{T.}~\bibnamefont{Tanaka}}, \bibinfo{author}{\bibfnamefont{K.}~\bibnamefont{Banerjee}}, \bibinfo{author}{\bibfnamefont{L.~T.} \bibnamefont{Bezzina}}, \bibinfo{author}{\bibfnamefont{I.~P.} \bibnamefont{Carter}}, \bibnamefont{et~al.}, \bibinfo{journal}{Physical Review C} \textbf{\bibinfo{volume}{102}}, \bibinfo{pages}{054611} (\bibinfo{year}{2020}).

\bibitem[{\citenamefont{{Swinton-Bland} et~al.}(2023)\citenamefont{{Swinton-Bland}, Buete, Hinde, Dasgupta, Tanaka, Berriman, Jeung, Banerjee, Bezzina, Carter et~al.}}]{Swinton-Bland2023a}
\bibinfo{author}{\bibfnamefont{B.~M.~A.} \bibnamefont{{Swinton-Bland}}}, \bibinfo{author}{\bibfnamefont{J.}~\bibnamefont{Buete}}, \bibinfo{author}{\bibfnamefont{D.~J.} \bibnamefont{Hinde}}, \bibinfo{author}{\bibfnamefont{M.}~\bibnamefont{Dasgupta}}, \bibinfo{author}{\bibfnamefont{T.}~\bibnamefont{Tanaka}}, \bibinfo{author}{\bibfnamefont{A.~C.} \bibnamefont{Berriman}}, \bibinfo{author}{\bibfnamefont{D.~Y.} \bibnamefont{Jeung}}, \bibinfo{author}{\bibfnamefont{K.}~\bibnamefont{Banerjee}}, \bibinfo{author}{\bibfnamefont{L.~T.} \bibnamefont{Bezzina}}, \bibinfo{author}{\bibfnamefont{I.~P.} \bibnamefont{Carter}}, \bibnamefont{et~al.}, \bibinfo{journal}{Physics Letters B} \textbf{\bibinfo{volume}{837}}, \bibinfo{pages}{137655} (\bibinfo{year}{2023}).

\bibitem[{\citenamefont{Banerjee et~al.}(2022)\citenamefont{Banerjee, Kozulin, Burtebayev, Gikal, Knyazheva, Itkis, Novikov, Kvochkina, Mukhamejanov, and Pan}}]{Banerjee2022}
\bibinfo{author}{\bibfnamefont{T.}~\bibnamefont{Banerjee}}, \bibinfo{author}{\bibfnamefont{E.~M.} \bibnamefont{Kozulin}}, \bibinfo{author}{\bibfnamefont{N.~T.} \bibnamefont{Burtebayev}}, \bibinfo{author}{\bibfnamefont{K.~B.} \bibnamefont{Gikal}}, \bibinfo{author}{\bibfnamefont{G.~N.} \bibnamefont{Knyazheva}}, \bibinfo{author}{\bibfnamefont{I.~M.} \bibnamefont{Itkis}}, \bibinfo{author}{\bibfnamefont{K.~V.} \bibnamefont{Novikov}}, \bibinfo{author}{\bibfnamefont{T.~N.} \bibnamefont{Kvochkina}}, \bibinfo{author}{\bibfnamefont{Y.~S.} \bibnamefont{Mukhamejanov}}, \bibnamefont{and} \bibinfo{author}{\bibfnamefont{A.~N.} \bibnamefont{Pan}}, \bibinfo{journal}{Physical Review C} \textbf{\bibinfo{volume}{105}}, \bibinfo{pages}{044614} (\bibinfo{year}{2022}).

\bibitem[{\citenamefont{Nag et~al.}(2021)\citenamefont{Nag, Tripathi, Patra, Mhatre, Santra, Rout, Kundu, Chattopadhyay, Pal, and Pujari}}]{Nag2021}
\bibinfo{author}{\bibfnamefont{T.~N.} \bibnamefont{Nag}}, \bibinfo{author}{\bibfnamefont{R.}~\bibnamefont{Tripathi}}, \bibinfo{author}{\bibfnamefont{S.}~\bibnamefont{Patra}}, \bibinfo{author}{\bibfnamefont{A.}~\bibnamefont{Mhatre}}, \bibinfo{author}{\bibfnamefont{S.}~\bibnamefont{Santra}}, \bibinfo{author}{\bibfnamefont{P.~C.} \bibnamefont{Rout}}, \bibinfo{author}{\bibfnamefont{A.}~\bibnamefont{Kundu}}, \bibinfo{author}{\bibfnamefont{D.}~\bibnamefont{Chattopadhyay}}, \bibinfo{author}{\bibfnamefont{A.}~\bibnamefont{Pal}}, \bibnamefont{and} \bibinfo{author}{\bibfnamefont{P.~K.} \bibnamefont{Pujari}}, \bibinfo{journal}{Physical Review C} \textbf{\bibinfo{volume}{103}}, \bibinfo{pages}{034612} (\bibinfo{year}{2021}).

\bibitem[{\citenamefont{Prasad et~al.}(2020)\citenamefont{Prasad, Hinde, Dasgupta, Jeung, Berriman, {Swinton-Bland}, Simenel, Simpson, Bernard, Williams et~al.}}]{Prasad2020a}
\bibinfo{author}{\bibfnamefont{E.}~\bibnamefont{Prasad}}, \bibinfo{author}{\bibfnamefont{D.~J.} \bibnamefont{Hinde}}, \bibinfo{author}{\bibfnamefont{M.}~\bibnamefont{Dasgupta}}, \bibinfo{author}{\bibfnamefont{D.~Y.} \bibnamefont{Jeung}}, \bibinfo{author}{\bibfnamefont{A.~C.} \bibnamefont{Berriman}}, \bibinfo{author}{\bibfnamefont{B.~M.~A.} \bibnamefont{{Swinton-Bland}}}, \bibinfo{author}{\bibfnamefont{C.}~\bibnamefont{Simenel}}, \bibinfo{author}{\bibfnamefont{E.~C.} \bibnamefont{Simpson}}, \bibinfo{author}{\bibfnamefont{R.}~\bibnamefont{Bernard}}, \bibinfo{author}{\bibfnamefont{E.}~\bibnamefont{Williams}}, \bibnamefont{et~al.}, \bibinfo{journal}{Physics Letters B} \textbf{\bibinfo{volume}{811}}, \bibinfo{pages}{135941} (\bibinfo{year}{2020}).

\bibitem[{\citenamefont{Berriman et~al.}(2022)\citenamefont{Berriman, Hinde, Jeung, Dasgupta, Haba, Tanaka, Banerjee, Banerjee, Bezzina, Buete et~al.}}]{Berriman2022}
\bibinfo{author}{\bibfnamefont{A.~C.} \bibnamefont{Berriman}}, \bibinfo{author}{\bibfnamefont{D.~J.} \bibnamefont{Hinde}}, \bibinfo{author}{\bibfnamefont{D.~Y.} \bibnamefont{Jeung}}, \bibinfo{author}{\bibfnamefont{M.}~\bibnamefont{Dasgupta}}, \bibinfo{author}{\bibfnamefont{H.}~\bibnamefont{Haba}}, \bibinfo{author}{\bibfnamefont{T.}~\bibnamefont{Tanaka}}, \bibinfo{author}{\bibfnamefont{K.}~\bibnamefont{Banerjee}}, \bibinfo{author}{\bibfnamefont{T.}~\bibnamefont{Banerjee}}, \bibinfo{author}{\bibfnamefont{L.~T.} \bibnamefont{Bezzina}}, \bibinfo{author}{\bibfnamefont{J.}~\bibnamefont{Buete}}, \bibnamefont{et~al.}, \bibinfo{journal}{Physical Review C} \textbf{\bibinfo{volume}{105}}, \bibinfo{pages}{064614} (\bibinfo{year}{2022}).

\bibitem[{\citenamefont{Goutte et~al.}(2005)\citenamefont{Goutte, Berger, Casoli, and Gogny}}]{Goutte2005}
\bibinfo{author}{\bibfnamefont{H.}~\bibnamefont{Goutte}}, \bibinfo{author}{\bibfnamefont{J.~F.} \bibnamefont{Berger}}, \bibinfo{author}{\bibfnamefont{P.}~\bibnamefont{Casoli}}, \bibnamefont{and} \bibinfo{author}{\bibfnamefont{D.}~\bibnamefont{Gogny}}, \bibinfo{journal}{Phys. Rev. C} \textbf{\bibinfo{volume}{71}}, \bibinfo{pages}{024316} (\bibinfo{year}{2005}).

\bibitem[{\citenamefont{Schunck and Robledo}(2016)}]{Schunck2016}
\bibinfo{author}{\bibfnamefont{N.}~\bibnamefont{Schunck}} \bibnamefont{and} \bibinfo{author}{\bibfnamefont{L.~M.} \bibnamefont{Robledo}}, \bibinfo{journal}{Reports on Progress in Physics} \textbf{\bibinfo{volume}{79}} (\bibinfo{year}{2016}), \eprint{1511.07517}.

\bibitem[{\citenamefont{Bartel et~al.}(2014)\citenamefont{Bartel, {Nerlo-Pomorska}, Pomorski, and Schmitt}}]{Bartel2014}
\bibinfo{author}{\bibfnamefont{J.}~\bibnamefont{Bartel}}, \bibinfo{author}{\bibfnamefont{B.}~\bibnamefont{{Nerlo-Pomorska}}}, \bibinfo{author}{\bibfnamefont{K.}~\bibnamefont{Pomorski}}, \bibnamefont{and} \bibinfo{author}{\bibfnamefont{C.}~\bibnamefont{Schmitt}}, \bibinfo{journal}{Physica Scripta} \textbf{\bibinfo{volume}{89}}, \bibinfo{pages}{054003} (\bibinfo{year}{2014}).

\bibitem[{\citenamefont{Randrup and M{\"o}ller}(2011)}]{Randrup2011b}
\bibinfo{author}{\bibfnamefont{J.}~\bibnamefont{Randrup}} \bibnamefont{and} \bibinfo{author}{\bibfnamefont{P.}~\bibnamefont{M{\"o}ller}}, \bibinfo{journal}{Physical Review Letters} \textbf{\bibinfo{volume}{106}}, \bibinfo{pages}{132503} (\bibinfo{year}{2011}).

\bibitem[{\citenamefont{McDonnell et~al.}(2014)\citenamefont{McDonnell, Nazarewicz, Sheikh, Staszczak, and Warda}}]{McDonnell2014}
\bibinfo{author}{\bibfnamefont{J.~D.} \bibnamefont{McDonnell}}, \bibinfo{author}{\bibfnamefont{W.}~\bibnamefont{Nazarewicz}}, \bibinfo{author}{\bibfnamefont{J.~A.} \bibnamefont{Sheikh}}, \bibinfo{author}{\bibfnamefont{A.}~\bibnamefont{Staszczak}}, \bibnamefont{and} \bibinfo{author}{\bibfnamefont{M.}~\bibnamefont{Warda}}, \bibinfo{journal}{Physical Review C - Nuclear Physics} \textbf{\bibinfo{volume}{90}}, \bibinfo{pages}{1} (\bibinfo{year}{2014}), \eprint{1406.6955}.

\bibitem[{\citenamefont{Schunck et~al.}(2014)\citenamefont{Schunck, Duke, Carr, and Knoll}}]{Schunck2014}
\bibinfo{author}{\bibfnamefont{N.}~\bibnamefont{Schunck}}, \bibinfo{author}{\bibfnamefont{D.}~\bibnamefont{Duke}}, \bibinfo{author}{\bibfnamefont{H.}~\bibnamefont{Carr}}, \bibnamefont{and} \bibinfo{author}{\bibfnamefont{A.}~\bibnamefont{Knoll}}, \bibinfo{journal}{Physical Review C - Nuclear Physics} \textbf{\bibinfo{volume}{90}} (\bibinfo{year}{2014}), \eprint{1311.2616}.

\bibitem[{\citenamefont{Kostryukov et~al.}(2021)\citenamefont{Kostryukov, Dobrowolski, {Nerlo-Pomorska}, Warda, Xiao, Chen, Liu, Tian, and Pomorski}}]{Kostryukov2021}
\bibinfo{author}{\bibfnamefont{P.~V.} \bibnamefont{Kostryukov}}, \bibinfo{author}{\bibfnamefont{A.}~\bibnamefont{Dobrowolski}}, \bibinfo{author}{\bibfnamefont{B.}~\bibnamefont{{Nerlo-Pomorska}}}, \bibinfo{author}{\bibfnamefont{M.}~\bibnamefont{Warda}}, \bibinfo{author}{\bibfnamefont{Z.~G.} \bibnamefont{Xiao}}, \bibinfo{author}{\bibfnamefont{Y.~J.} \bibnamefont{Chen}}, \bibinfo{author}{\bibfnamefont{L.~L.} \bibnamefont{Liu}}, \bibinfo{author}{\bibfnamefont{J.~L.} \bibnamefont{Tian}}, \bibnamefont{and} \bibinfo{author}{\bibfnamefont{K.}~\bibnamefont{Pomorski}}, \bibinfo{journal}{Chinese Physics C}  (\bibinfo{year}{2021}), \eprint{2107.09981}.

\bibitem[{\citenamefont{Zdeb et~al.}(2021)\citenamefont{Zdeb, Warda, and Robledo}}]{Zdeb2021}
\bibinfo{author}{\bibfnamefont{A.}~\bibnamefont{Zdeb}}, \bibinfo{author}{\bibfnamefont{M.}~\bibnamefont{Warda}}, \bibnamefont{and} \bibinfo{author}{\bibfnamefont{L.~M.} \bibnamefont{Robledo}}, \bibinfo{journal}{Physical Review C} \textbf{\bibinfo{volume}{104}} (\bibinfo{year}{2021}).

\bibitem[{\citenamefont{Okada et~al.}(2023)\citenamefont{Okada, Wada, Capote, and Carjan}}]{Okada2023a}
\bibinfo{author}{\bibfnamefont{K.}~\bibnamefont{Okada}}, \bibinfo{author}{\bibfnamefont{T.}~\bibnamefont{Wada}}, \bibinfo{author}{\bibfnamefont{R.}~\bibnamefont{Capote}}, \bibnamefont{and} \bibinfo{author}{\bibfnamefont{N.}~\bibnamefont{Carjan}}, \bibinfo{journal}{Physical Review C} \textbf{\bibinfo{volume}{107}}, \bibinfo{pages}{034608} (\bibinfo{year}{2023}).

\bibitem[{\citenamefont{Pa{\c s}ca et~al.}(2016)\citenamefont{Pa{\c s}ca, Andreev, Adamian, and Antonenko}}]{Pasca2016}
\bibinfo{author}{\bibfnamefont{H.}~\bibnamefont{Pa{\c s}ca}}, \bibinfo{author}{\bibfnamefont{A.~V.} \bibnamefont{Andreev}}, \bibinfo{author}{\bibfnamefont{G.~G.} \bibnamefont{Adamian}}, \bibnamefont{and} \bibinfo{author}{\bibfnamefont{N.~V.} \bibnamefont{Antonenko}}, \bibinfo{journal}{Physics Letters B} \textbf{\bibinfo{volume}{760}}, \bibinfo{pages}{800} (\bibinfo{year}{2016}).

\bibitem[{\citenamefont{Pa{\c s}ca et~al.}(2018)\citenamefont{Pa{\c s}ca, Andreev, Adamian, and Antonenko}}]{Pasca2018}
\bibinfo{author}{\bibfnamefont{H.}~\bibnamefont{Pa{\c s}ca}}, \bibinfo{author}{\bibfnamefont{A.~V.} \bibnamefont{Andreev}}, \bibinfo{author}{\bibfnamefont{G.~G.} \bibnamefont{Adamian}}, \bibnamefont{and} \bibinfo{author}{\bibfnamefont{N.~V.} \bibnamefont{Antonenko}}, \bibinfo{journal}{Physical Review C} \textbf{\bibinfo{volume}{97}}, \bibinfo{pages}{034621} (\bibinfo{year}{2018}).

\bibitem[{\citenamefont{Pa{\c s}ca et~al.}(2020)\citenamefont{Pa{\c s}ca, Andreev, Adamian, and Antonenko}}]{Pasca2020}
\bibinfo{author}{\bibfnamefont{H.}~\bibnamefont{Pa{\c s}ca}}, \bibinfo{author}{\bibfnamefont{A.~V.} \bibnamefont{Andreev}}, \bibinfo{author}{\bibfnamefont{G.~G.} \bibnamefont{Adamian}}, \bibnamefont{and} \bibinfo{author}{\bibfnamefont{N.~V.} \bibnamefont{Antonenko}}, \bibinfo{journal}{Physical Review C} \textbf{\bibinfo{volume}{101}}, \bibinfo{pages}{064604} (\bibinfo{year}{2020}).

\bibitem[{\citenamefont{Pa{\c s}ca et~al.}(2021)\citenamefont{Pa{\c s}ca, Andreev, Adamian, and Antonenko}}]{Pasca2021}
\bibinfo{author}{\bibfnamefont{H.}~\bibnamefont{Pa{\c s}ca}}, \bibinfo{author}{\bibfnamefont{A.~V.} \bibnamefont{Andreev}}, \bibinfo{author}{\bibfnamefont{G.~G.} \bibnamefont{Adamian}}, \bibnamefont{and} \bibinfo{author}{\bibfnamefont{N.~V.} \bibnamefont{Antonenko}}, \bibinfo{journal}{Physical Review C} \textbf{\bibinfo{volume}{104}}, \bibinfo{pages}{014604} (\bibinfo{year}{2021}).

\bibitem[{\citenamefont{Carjan et~al.}(2017)\citenamefont{Carjan, Ivanyuk, and Oganessian}}]{Carjan2017}
\bibinfo{author}{\bibfnamefont{N.}~\bibnamefont{Carjan}}, \bibinfo{author}{\bibfnamefont{F.~A.} \bibnamefont{Ivanyuk}}, \bibnamefont{and} \bibinfo{author}{\bibfnamefont{Y.~T.} \bibnamefont{Oganessian}}, \bibinfo{journal}{Nuclear Physics A} \textbf{\bibinfo{volume}{968}}, \bibinfo{pages}{453} (\bibinfo{year}{2017}).

\bibitem[{\citenamefont{Carjan et~al.}(2019)\citenamefont{Carjan, Ivanyuk, and Oganessian}}]{Carjan2019}
\bibinfo{author}{\bibfnamefont{N.}~\bibnamefont{Carjan}}, \bibinfo{author}{\bibfnamefont{F.~A.} \bibnamefont{Ivanyuk}}, \bibnamefont{and} \bibinfo{author}{\bibfnamefont{Y.~T.} \bibnamefont{Oganessian}}, \bibinfo{journal}{Physical Review C} \textbf{\bibinfo{volume}{99}} (\bibinfo{year}{2019}), \eprint{1811.09913}.

\bibitem[{\citenamefont{Carjan et~al.}(2015)\citenamefont{Carjan, Ivanyuk, Oganessian, and {Ter-Akopian}}}]{Carjan2015}
\bibinfo{author}{\bibfnamefont{N.}~\bibnamefont{Carjan}}, \bibinfo{author}{\bibfnamefont{F.}~\bibnamefont{Ivanyuk}}, \bibinfo{author}{\bibfnamefont{{\relax Yu}.}~\bibnamefont{Oganessian}}, \bibnamefont{and} \bibinfo{author}{\bibfnamefont{G.}~\bibnamefont{{Ter-Akopian}}}, \bibinfo{journal}{Nuclear Physics A} \textbf{\bibinfo{volume}{942}}, \bibinfo{pages}{97} (\bibinfo{year}{2015}).

\bibitem[{\citenamefont{Pa{\c s}ca et~al.}(2023)\citenamefont{Pa{\c s}ca, Andreev, Adamian, and Antonenko}}]{Pasca2023}
\bibinfo{author}{\bibfnamefont{H.}~\bibnamefont{Pa{\c s}ca}}, \bibinfo{author}{\bibfnamefont{A.~V.} \bibnamefont{Andreev}}, \bibinfo{author}{\bibfnamefont{G.~G.} \bibnamefont{Adamian}}, \bibnamefont{and} \bibinfo{author}{\bibfnamefont{N.~V.} \bibnamefont{Antonenko}}, \bibinfo{journal}{Physical Review C} \textbf{\bibinfo{volume}{108}}, \bibinfo{pages}{014613} (\bibinfo{year}{2023}).

\bibitem[{\citenamefont{M{\"o}ller and Randrup}(2015)}]{Moller2015a}
\bibinfo{author}{\bibfnamefont{P.}~\bibnamefont{M{\"o}ller}} \bibnamefont{and} \bibinfo{author}{\bibfnamefont{J.}~\bibnamefont{Randrup}}, \bibinfo{journal}{Physical Review C} \textbf{\bibinfo{volume}{91}}, \bibinfo{pages}{044316} (\bibinfo{year}{2015}).

\bibitem[{\citenamefont{M{\"o}ller et~al.}(2012)\citenamefont{M{\"o}ller, Randrup, and Sierk}}]{Moller2012}
\bibinfo{author}{\bibfnamefont{P.}~\bibnamefont{M{\"o}ller}}, \bibinfo{author}{\bibfnamefont{J.}~\bibnamefont{Randrup}}, \bibnamefont{and} \bibinfo{author}{\bibfnamefont{A.~J.} \bibnamefont{Sierk}}, \bibinfo{journal}{Physical Review C} \textbf{\bibinfo{volume}{85}}, \bibinfo{pages}{024306} (\bibinfo{year}{2012}).

\bibitem[{\citenamefont{Aritomo and Chiba}(2013)}]{Aritomo2013}
\bibinfo{author}{\bibfnamefont{Y.}~\bibnamefont{Aritomo}} \bibnamefont{and} \bibinfo{author}{\bibfnamefont{S.}~\bibnamefont{Chiba}}, \bibinfo{journal}{Physical Review C} \textbf{\bibinfo{volume}{88}}, \bibinfo{pages}{044614} (\bibinfo{year}{2013}).

\bibitem[{\citenamefont{Aritomo et~al.}(2014)\citenamefont{Aritomo, Chiba, and Ivanyuk}}]{Aritomo2014}
\bibinfo{author}{\bibfnamefont{Y.}~\bibnamefont{Aritomo}}, \bibinfo{author}{\bibfnamefont{S.}~\bibnamefont{Chiba}}, \bibnamefont{and} \bibinfo{author}{\bibfnamefont{F.}~\bibnamefont{Ivanyuk}}, \bibinfo{journal}{Physical Review C} \textbf{\bibinfo{volume}{90}}, \bibinfo{pages}{054609} (\bibinfo{year}{2014}).

\bibitem[{\citenamefont{Liu et~al.}(2021)\citenamefont{Liu, Chen, Wu, Li, Ge, and Pomorski}}]{Liu2021}
\bibinfo{author}{\bibfnamefont{L.-L.} \bibnamefont{Liu}}, \bibinfo{author}{\bibfnamefont{Y.-J.} \bibnamefont{Chen}}, \bibinfo{author}{\bibfnamefont{X.-Z.} \bibnamefont{Wu}}, \bibinfo{author}{\bibfnamefont{Z.-X.} \bibnamefont{Li}}, \bibinfo{author}{\bibfnamefont{Z.-G.} \bibnamefont{Ge}}, \bibnamefont{and} \bibinfo{author}{\bibfnamefont{K.}~\bibnamefont{Pomorski}}, \bibinfo{journal}{Physical Review C} \textbf{\bibinfo{volume}{103}}, \bibinfo{pages}{044601} (\bibinfo{year}{2021}).

\bibitem[{\citenamefont{Schmitt et~al.}(2019)\citenamefont{Schmitt, Mazurek, and Nadtochy}}]{Schmitt2019}
\bibinfo{author}{\bibfnamefont{C.}~\bibnamefont{Schmitt}}, \bibinfo{author}{\bibfnamefont{K.}~\bibnamefont{Mazurek}}, \bibnamefont{and} \bibinfo{author}{\bibfnamefont{P.~N.} \bibnamefont{Nadtochy}}, \bibinfo{journal}{Physical Review C} \textbf{\bibinfo{volume}{100}}, \bibinfo{pages}{064606} (\bibinfo{year}{2019}).

\bibitem[{\citenamefont{Regnier et~al.}(2019)\citenamefont{Regnier, Dubray, and Schunck}}]{Regnier2019}
\bibinfo{author}{\bibfnamefont{D.}~\bibnamefont{Regnier}}, \bibinfo{author}{\bibfnamefont{N.}~\bibnamefont{Dubray}}, \bibnamefont{and} \bibinfo{author}{\bibfnamefont{N.}~\bibnamefont{Schunck}}, \bibinfo{journal}{Physical Review C} \textbf{\bibinfo{volume}{99}}, \bibinfo{pages}{024611} (\bibinfo{year}{2019}).

\bibitem[{\citenamefont{Verriere and Regnier}(2020)}]{Verriere2020a}
\bibinfo{author}{\bibfnamefont{M.}~\bibnamefont{Verriere}} \bibnamefont{and} \bibinfo{author}{\bibfnamefont{D.}~\bibnamefont{Regnier}}, \bibinfo{journal}{Frontiers in Physics} \textbf{\bibinfo{volume}{8}} (\bibinfo{year}{2020}), \eprint{2004.10147}.

\bibitem[{\citenamefont{Zhao et~al.}(2022)\citenamefont{Zhao, Nik{\v s}i{\'c}, and Vretenar}}]{Zhao2022}
\bibinfo{author}{\bibfnamefont{J.}~\bibnamefont{Zhao}}, \bibinfo{author}{\bibfnamefont{T.}~\bibnamefont{Nik{\v s}i{\'c}}}, \bibnamefont{and} \bibinfo{author}{\bibfnamefont{D.}~\bibnamefont{Vretenar}}, \bibinfo{journal}{Physical Review C} \textbf{\bibinfo{volume}{105}}, \bibinfo{pages}{1} (\bibinfo{year}{2022}), \eprint{2202.13637}.

\bibitem[{\citenamefont{Younes et~al.}(2019)\citenamefont{Younes, Gogny, and Berger}}]{Younes2019}
\bibinfo{author}{\bibfnamefont{W.}~\bibnamefont{Younes}}, \bibinfo{author}{\bibfnamefont{D.~M.} \bibnamefont{Gogny}}, \bibnamefont{and} \bibinfo{author}{\bibfnamefont{J.-F.} \bibnamefont{Berger}}, \emph{\bibinfo{title}{A {{Microscopic Theory}} of {{Fission Dynamics Based}} on the {{Generator Coordinate Method}}}}, vol. \bibinfo{volume}{950} (\bibinfo{publisher}{Springer}, \bibinfo{year}{2019}), ISBN \bibinfo{isbn}{978-3-030-04422-0}.

\bibitem[{\citenamefont{Nishio et~al.}(2015)\citenamefont{Nishio, Andreyev, Chapman, Derkx, Düllmann, Ghys, Heßberger, Hirose, Ikezoe, Khuyagbaatar et~al.}}]{Nishio2015}
\bibinfo{author}{\bibfnamefont{K.}~\bibnamefont{Nishio}}, \bibinfo{author}{\bibfnamefont{A.}~\bibnamefont{Andreyev}}, \bibinfo{author}{\bibfnamefont{R.}~\bibnamefont{Chapman}}, \bibinfo{author}{\bibfnamefont{X.}~\bibnamefont{Derkx}}, \bibinfo{author}{\bibfnamefont{C.}~\bibnamefont{Düllmann}}, \bibinfo{author}{\bibfnamefont{L.}~\bibnamefont{Ghys}}, \bibinfo{author}{\bibfnamefont{F.}~\bibnamefont{Heßberger}}, \bibinfo{author}{\bibfnamefont{K.}~\bibnamefont{Hirose}}, \bibinfo{author}{\bibfnamefont{H.}~\bibnamefont{Ikezoe}}, \bibinfo{author}{\bibfnamefont{J.}~\bibnamefont{Khuyagbaatar}}, \bibnamefont{et~al.}, \bibinfo{journal}{Phys. Lett. B} \textbf{\bibinfo{volume}{748}}, \bibinfo{pages}{89} (\bibinfo{year}{2015}).

\bibitem[{\citenamefont{Andreyev et~al.}(2010)\citenamefont{Andreyev, Elseviers, Huyse, Van~Duppen, Antalic, Barzakh, Bree, Cocolios, Comas, Diriken et~al.}}]{Andreyev2010}
\bibinfo{author}{\bibfnamefont{A.~N.} \bibnamefont{Andreyev}}, \bibinfo{author}{\bibfnamefont{J.}~\bibnamefont{Elseviers}}, \bibinfo{author}{\bibfnamefont{M.}~\bibnamefont{Huyse}}, \bibinfo{author}{\bibfnamefont{P.}~\bibnamefont{Van~Duppen}}, \bibinfo{author}{\bibfnamefont{S.}~\bibnamefont{Antalic}}, \bibinfo{author}{\bibfnamefont{A.}~\bibnamefont{Barzakh}}, \bibinfo{author}{\bibfnamefont{N.}~\bibnamefont{Bree}}, \bibinfo{author}{\bibfnamefont{T.~E.} \bibnamefont{Cocolios}}, \bibinfo{author}{\bibfnamefont{V.~F.} \bibnamefont{Comas}}, \bibinfo{author}{\bibfnamefont{J.}~\bibnamefont{Diriken}}, \bibnamefont{et~al.}, \bibinfo{journal}{Physical Review Letters} \textbf{\bibinfo{volume}{105}}, \bibinfo{pages}{252502} (\bibinfo{year}{2010}).

\bibitem[{\citenamefont{Prasad et~al.}(2015)\citenamefont{Prasad, Hinde, Ramachandran, Williams, Dasgupta, Carter, Cook, Jeung, Luong, McNeil et~al.}}]{prasad2015}
\bibinfo{author}{\bibfnamefont{E.}~\bibnamefont{Prasad}}, \bibinfo{author}{\bibfnamefont{D.~J.} \bibnamefont{Hinde}}, \bibinfo{author}{\bibfnamefont{K.}~\bibnamefont{Ramachandran}}, \bibinfo{author}{\bibfnamefont{E.}~\bibnamefont{Williams}}, \bibinfo{author}{\bibfnamefont{M.}~\bibnamefont{Dasgupta}}, \bibinfo{author}{\bibfnamefont{I.~P.} \bibnamefont{Carter}}, \bibinfo{author}{\bibfnamefont{K.~J.} \bibnamefont{Cook}}, \bibinfo{author}{\bibfnamefont{D.~Y.} \bibnamefont{Jeung}}, \bibinfo{author}{\bibfnamefont{D.~H.} \bibnamefont{Luong}}, \bibinfo{author}{\bibfnamefont{S.}~\bibnamefont{McNeil}}, \bibnamefont{et~al.}, \bibinfo{journal}{Phys. Rev. C} \textbf{\bibinfo{volume}{91}}, \bibinfo{pages}{064605} (\bibinfo{year}{2015}), \urlprefix\url{https://link.aps.org/doi/10.1103/PhysRevC.91.064605}.

\bibitem[{\citenamefont{Ichikawa et~al.}(2012)\citenamefont{Ichikawa, Iwamoto, M\"oller, and Sierk}}]{ichikawa2012}
\bibinfo{author}{\bibfnamefont{T.}~\bibnamefont{Ichikawa}}, \bibinfo{author}{\bibfnamefont{A.}~\bibnamefont{Iwamoto}}, \bibinfo{author}{\bibfnamefont{P.}~\bibnamefont{M\"oller}}, \bibnamefont{and} \bibinfo{author}{\bibfnamefont{A.~J.} \bibnamefont{Sierk}}, \bibinfo{journal}{Phys. Rev. C} \textbf{\bibinfo{volume}{86}}, \bibinfo{pages}{024610} (\bibinfo{year}{2012}), \urlprefix\url{https://link.aps.org/doi/10.1103/PhysRevC.86.024610}.

\bibitem[{\citenamefont{Andreev et~al.}(2012)\citenamefont{Andreev, Adamian, and Antonenko}}]{Andreev2012}
\bibinfo{author}{\bibfnamefont{A.~V.} \bibnamefont{Andreev}}, \bibinfo{author}{\bibfnamefont{G.~G.} \bibnamefont{Adamian}}, \bibnamefont{and} \bibinfo{author}{\bibfnamefont{N.~V.} \bibnamefont{Antonenko}}, \bibinfo{journal}{Physical Review C} \textbf{\bibinfo{volume}{86}}, \bibinfo{pages}{044315} (\bibinfo{year}{2012}).

\bibitem[{\citenamefont{Andreev et~al.}(2013)\citenamefont{Andreev, Adamian, Antonenko, and Andreyev}}]{andreev2013}
\bibinfo{author}{\bibfnamefont{A.~V.} \bibnamefont{Andreev}}, \bibinfo{author}{\bibfnamefont{G.~G.} \bibnamefont{Adamian}}, \bibinfo{author}{\bibfnamefont{N.~V.} \bibnamefont{Antonenko}}, \bibnamefont{and} \bibinfo{author}{\bibfnamefont{A.~N.} \bibnamefont{Andreyev}}, \bibinfo{journal}{Phys. Rev. C} \textbf{\bibinfo{volume}{88}}, \bibinfo{pages}{047604} (\bibinfo{year}{2013}), \urlprefix\url{https://link.aps.org/doi/10.1103/PhysRevC.88.047604}.

\bibitem[{\citenamefont{Panebianco et~al.}(2012)\citenamefont{Panebianco, Sida, Goutte, Lema\^{\i}tre, Dubray, and Hilaire}}]{panebianco2012}
\bibinfo{author}{\bibfnamefont{S.}~\bibnamefont{Panebianco}}, \bibinfo{author}{\bibfnamefont{J.-L.} \bibnamefont{Sida}}, \bibinfo{author}{\bibfnamefont{H.}~\bibnamefont{Goutte}}, \bibinfo{author}{\bibfnamefont{J.-F. m.~c.} \bibnamefont{Lema\^{\i}tre}}, \bibinfo{author}{\bibfnamefont{N.}~\bibnamefont{Dubray}}, \bibnamefont{and} \bibinfo{author}{\bibfnamefont{S.}~\bibnamefont{Hilaire}}, \bibinfo{journal}{Phys. Rev. C} \textbf{\bibinfo{volume}{86}}, \bibinfo{pages}{064601} (\bibinfo{year}{2012}), \urlprefix\url{https://link.aps.org/doi/10.1103/PhysRevC.86.064601}.

\bibitem[{\citenamefont{Warda et~al.}(2012)\citenamefont{Warda, Staszczak, and Nazarewicz}}]{warda2012}
\bibinfo{author}{\bibfnamefont{M.}~\bibnamefont{Warda}}, \bibinfo{author}{\bibfnamefont{A.}~\bibnamefont{Staszczak}}, \bibnamefont{and} \bibinfo{author}{\bibfnamefont{W.}~\bibnamefont{Nazarewicz}}, \bibinfo{journal}{Phys. Rev. C} \textbf{\bibinfo{volume}{86}}, \bibinfo{pages}{024601} (\bibinfo{year}{2012}), \urlprefix\url{https://link.aps.org/doi/10.1103/PhysRevC.86.024601}.

\bibitem[{\citenamefont{{Bernard, Rémi N.} et~al.}(2024)\citenamefont{{Bernard, Rémi N.}, {Simenel, Cédric}, {Blanchon, Guillaume}, {Lau, Ngee-Wein T.}, and {McGlynn, Patrick}}}]{bernard2024}
\bibinfo{author}{\bibnamefont{{Bernard, Rémi N.}}}, \bibinfo{author}{\bibnamefont{{Simenel, Cédric}}}, \bibinfo{author}{\bibnamefont{{Blanchon, Guillaume}}}, \bibinfo{author}{\bibnamefont{{Lau, Ngee-Wein T.}}}, \bibnamefont{and} \bibinfo{author}{\bibnamefont{{McGlynn, Patrick}}}, \bibinfo{journal}{Eur. Phys. J. A} \textbf{\bibinfo{volume}{60}}, \bibinfo{pages}{192} (\bibinfo{year}{2024}), \urlprefix\url{https://doi.org/10.1140/epja/s10050-024-01415-2}.

\bibitem[{\citenamefont{B\"{o}ckstiegel et~al.}(2008)\citenamefont{B\"{o}ckstiegel, Steinh\"{a}user, Schmidt, Clerc, Grewe, Heinz, {de Jong}, Junghans, M\"{u}ller, and Voss}}]{bockstiegel2008}
\bibinfo{author}{\bibfnamefont{C.}~\bibnamefont{B\"{o}ckstiegel}}, \bibinfo{author}{\bibfnamefont{S.}~\bibnamefont{Steinh\"{a}user}}, \bibinfo{author}{\bibfnamefont{K.-H.} \bibnamefont{Schmidt}}, \bibinfo{author}{\bibfnamefont{H.-G.} \bibnamefont{Clerc}}, \bibinfo{author}{\bibfnamefont{A.}~\bibnamefont{Grewe}}, \bibinfo{author}{\bibfnamefont{A.}~\bibnamefont{Heinz}}, \bibinfo{author}{\bibfnamefont{M.}~\bibnamefont{{de Jong}}}, \bibinfo{author}{\bibfnamefont{A.~R.} \bibnamefont{Junghans}}, \bibinfo{author}{\bibfnamefont{J.}~\bibnamefont{M\"{u}ller}}, \bibnamefont{and} \bibinfo{author}{\bibfnamefont{B.}~\bibnamefont{Voss}}, \bibinfo{journal}{Nucl. Phys. A} \textbf{\bibinfo{volume}{802}}, \bibinfo{pages}{12} (\bibinfo{year}{2008}).

\bibitem[{\citenamefont{Bonneau}(2006)}]{bonneau2006}
\bibinfo{author}{\bibfnamefont{L.}~\bibnamefont{Bonneau}}, \bibinfo{journal}{Phys. Rev. C} \textbf{\bibinfo{volume}{74}}, \bibinfo{pages}{014301} (\bibinfo{year}{2006}).

\end{thebibliography}
  
	\end{document}